**On the Elastic Anisotropy of the Entropy-Stabilized Oxide (Mg, Co, Ni, Cu, Zn)O**

Krishna Chaitanya Pitike[1], Andres Marquez-Rossy[1], Alexis Flores-Betancourt[1], De Xin Chen[1], Santosh KC[2], Valentino R. Cooper[1] and Edgar Lara-Curzio[1]*

1 Materials Science & Technology Division, Oak Ridge National Laboratory, 1 Bethel Valley Rd., Oak Ridge, Tennessee 37831, USA

2 Chemical & Materials Engineering, San José State University, San José, California 95112, USA

* Corresponding author

E-mail: laracurzioe@ornl.gov

**Abstract**

In this paper, we study the elastic properties of the entropy-stabilized oxide (Mg, Co, Ni, Cu, Zn)O, using experimental and first principles techniques. Our measurements of the indentation modulus on grains with a wide range of crystallographic orientations of the entropy-stabilized oxide revealed a high degree of elastic isotropy at ambient conditions. First principles calculations predict mild elastic anisotropy for the paramagnetic structure, which decreases when the system is considered to be non-magnetic. When the antiferromagnetic state of CoO, CuO and NiO is accounted for in the calculations, a slight increase in the elastic anisotropy is observed, suggesting a coupling between magnetic ordering and the orientation dependent elastic properties. Furthermore, an examination of the local structure reveals that the isotropy is favored through local ionic distortions of Cu and Zn – due to their tendency to form tenorite and wurtzite phases. The relationships between the elastic properties of the multicomponent oxide and those of its constituent binary oxides are reviewed. These insights open up new avenues for controlling isotropy for technological applications through tuning composition and structure in the entropy-stabilized oxide or the high entropy compounds in general.

## I. INTRODUCTION

Elastic anisotropy is known to affect phonon modes,[1] phase transformations,[2] dislocation dynamics,[3] crack propogation,[4] charge defect mobility,[5] among many others. Hence elastic anisotropy has been a property of interest to the research community studying mechanical and related physical properties of materials. Understanding the influence of mixing several chemical species – for example a high entropy compound – on the elastic anisotropy is of paramount importance for engineering new high entropy materials for mechanical applications.

Since the discovery of a multicomponent entropy stabilized oxide in the rock salt phase,[6] there has been a surge in the design and synthesis of entropy-stabilized compounds. These include, perovskite oxides,[7–9] fluorites,[10,11] spinel,[12] boride[13] and carbides[14–18]. Early reports indicate many interesting emergent properties, such as, long range magnetic ordering,[19] reversible energy storage,[20] and giant exchange-coupling enhancement[21]. However, the role of chemical disorder on the physical and chemical properties of the entropy stabilized material is not well understood. Another key question is, how does the behavior of an entropy-stabilized material differ from that of its end-member components? Here, we explore the linear elastic properties of the Entropy-Stabilized Oxide – (Mg, Co, Ni, Cu, Zn)O, henceforth denoted as simply ESO – stabilized in the rock salt phase. Specifically, we report on the anisotropy in elastic properties of the ESO using both first principles computations and experimental measurements.

The first work to address the importance of quantifying anisotropy was by Zener[22] in 1930. In his work, Zener proposed the following parameter to quantify the anisotropy of cubic crystals[22] :

$$A^{\mathrm{Z}} = \frac{2\,(s_{11}-s_{12})}{s_{44}} = \frac{2\,c_{44}}{c_{11}-c_{12}} \tag{1}$$

where $s_{ij}$ and $c_{ij}$ refer to the compliance and stiffness constants of the material, respectively. Zener's anisotropy factor represents the ratio of the two extreme elastic shear coefficients in a cubic crystal: $c_{44}$ representing resistance to shear on $\{100\}$ in the $\langle 0kl \rangle$ direction and $(c_{11}-c_{12})/2$ representing resistance to shear on $\{110\}$ in the $\langle 110 \rangle$ direction.[22] When a cubic crystal is isotropic, $A^{\mathrm{Z}} = 1$ and deviations from unity are indicative of the degree of anisotropy.

In most cases $A^{\mathrm{Z}} > 1$, but there are bcc crystals, such as polonium and some oxide fluorites and halides, for which $A^{\mathrm{Z}} < 1$.[23,24]

Later, Ranganathan and Ostoja-Starzewski derived an elastic anisotropy index $A^{\mathrm{U}}$ to provide a measure of elastic anisotropy regardless of the symmetry of the crystal.[25] This parameter is based on the fractional difference between the upper and lower bounds on the bulk and shear moduli according to the following relationship:

$$A^{\mathrm{U}} = 5\frac{G^{\mathrm{V}}}{G^{\mathrm{R}}} + \frac{K^{\mathrm{V}}}{K^{\mathrm{R}}} - 6 \qquad (2)$$

where $G$ and $K$ are the shear and bulk modulus, respectively and $\mathrm{V}$ and $\mathrm{R}$ correspond to the Voigt (upper bound) and the Reuss (lower bound) estimates, respectively. For isotropic crystals, $A^{\mathrm{U}}$ is equal to zero and the extent of anisotropy is defined by the departure of $A^{\mathrm{U}}$ from zero.[25]

Here, we show that the ESO exhibits a high degree of elastic isotropy through indentation modulus measurements and first principles estimated elastic properties. We find that the elastic properties, such as the anisotropy index, bulk and shear moduli, are well represented by a linear combination of the respective properties of the parent binary oxides. Through first principles calculations, we probe the influence of the local structure and magnetic configuration on the elastic anisotropy. Specifically, we find that the local distortions around Cu and Zn cations in the nonmagnetic structure – as a result of their tendency to form tenorite and wurtzite phases – promote isotropy. However, we find increasing degrees of elastic anisotropy in the paramagnetic and antiferromagnetic phases due to the suppressed local distortions and magnetic anisotropy in the antiferromagnetic structure. Antiferromagnetic ordering is only observed below the Néel temperature $T_{\mathrm{N}} = 113$ K,[19] suggesting a strong temperature dependence of the elastic isotropy due to a magnetic phase transition. Similarly, the mixed phase to single phase transition at high temperature may also induce reduction in anisotropy at higher temperature. Together these results present a large phase space through which the elastic properties of ESO may be tuned and thus have implications for its use in different applications.

## II. EXPERIMENTAL

### *Sample*

The material used in this investigation was obtained by consolidating mixed oxide powders prepared by the polymerized organic-inorganic route[26]. The powders were hot-pressed at 1,100°C for three hours under a compressive stress of 55 MPa.

The chemical composition, microstructure, crystal structure and crystal orientation of the consolidated ESO material were determined by scanning electron microscopy, electron dispersive spectroscopy, X-ray diffraction and electron backscatter diffraction (EBSD). Details about the characterization of the material are provided as supplemental material.

### *Indentation*

Several arrays of $20 \times 20$ nanoindents were made on grains with different crystallographic orientations using a nanoindenter equipped with a Berkovich diamond indenter. The indentation modulus, $M$, was calculated from the experimentally-determined reduced modulus, $M_r$, according to the following relationship[27],

$$\frac{1}{M} = \frac{1}{M_r} - \left(\frac{1-\nu_i^2}{E_i}\right) \quad (3)$$

Where $\nu_i = 0.07$ and $E_i = 1{,}141$ GPa are the Poisson's ratio and Young's modulus of the diamond indenter, respectively (TI 950 TriboIndenter User Manual Revision 9.3.0314, Hysitron Incorporated (2014)).

The indentation modulus $M(hkl)$ of the $(hkl)$ surface of a single crystal can be predicted as the product of a correction factor $\beta_{hkl}$ and the indentation modulus of an isotropic, randomly oriented, polycrystalline aggregate of the same material[27],

$$M_{hkl} = \alpha\, \beta_{hkl} \left(\frac{E}{1-\nu^2}\right)_{isotropic} \quad (4)$$

where $\alpha$ is an additional correction factor to account for the geometry of the indenter tip and its orientation with respect to the crystallographic orientation of the indented surface. The correction factor $\beta_{hkl}$ for crystals with cubic symmetry is given by:

$$\beta_{hkl} = a + c(Z - A_0)^B \quad (5)$$

where $a$, $c$, $A_0$, $and$ $B$ are a function of Poisson's ratio in the cube directions.[27] The indentation modulus of the isotropic, randomly oriented, polycrystalline aggregate can be determined using the Voigt-Reuss-Hill average which are calculated using the elastic constants of a single crystal.[28] Additional details are presented as supplemental material.

### *Theoretical and Computational Details*

First-principles calculations were performed using density functional theory (DFT) as implemented in the Vienna Ab initio Simulation Package (VASP)[29,30] with the Perdew-Burke-Ernzerhof for solids (PBEsol) spin density approximation[31] for exchange correlation. The projector-augmented plane-wave method[32,33] with a 600 eV energy cutoff was used. For a primitive unit cell in the rock salt phase containing 2 atoms, a zone-edge-shifted $8 \times 8 \times 8$ automatically generated k-point mesh was used for the Brillouin zone (BZ) integrations. For the larger supercells required for modelling the ESO – containing 160 atoms – a $\Gamma$ point calculation was done. In all DFT calculations, internal ionic positions were relaxed to forces less than 0.005 eV/Å. The lattice constants were relaxed until the appropriate stress tensor components were less than 0.1 kbar.

To account for the Coulomb interactions within the partially filled $d$–orbitals of the Co, Ni and Cu cations in their respective binary oxides and also the ESO, we employ the DFT$+U$ approach using the simplified (rotationally invariant) approach.[34] $U = 6$ eV was used for Co, Ni and Cu $d$–states, in order to find the stable ground phases of their respective binary oxides – rock salt phase for CoO, NiO and Tenorite phase for CuO. Furthermore, CoO, NiO and the AFM model of ESO were initialized to an antiferromagnetic (AFM) ordering state with ferromagnetic planes parallel to (111) planes.

Figure 1(a) and (b) shows the rock salt structure in the $Fm\bar{3}m$ space group with 4 and 1 $A$O formula units per unit cells, respectively. The primitive unit cell (PC) was obtained from the FCC unit cell by the rotation matrix shown below the arrow. The ESO is modelled using a $4 \times 4 \times 5$ supercell of PC where $A$ = 0.2×{Mg, Co, Ni Cu, Zn}. The five cations are quasi-randomly assigned to the $A$-site using the special quasi-random structures algorithm introduced by Zunger et. al.[35]

and implemented in Alloy Theoretic Automated Toolkit (ATAT) open source software[36]. The lattice constants for the constituent binary oxide and ESO models in their respective ground phases are compared with experimental values in Table SIII of supplemental material. Moreover, the lattice constants for CuO and ZnO are also provided in the cubic phase.

The linear elastic stiffness coefficients were calculated from DFT using a method similar to that used for calculating the elastic constants presented in the Materials Project (MP).[37,38] Details can be found as supplemental material. Values of the Néel temperature, $T_{\mathrm{N}}$ of CoO, NiO, CuO and ESO have been experimentally determined as 291 K, 525 K, 220K and 113K, respectively.[19,39,40] The predicted values of Young's modulus of the ESO in this work are compared with experimental values determined at room temperature, $\mathrm{RT}{\sim}300\ \mathrm{K}$. Therefore, since the value of $T_{\mathrm{N}}$ for the ESO compound, $T_{\mathrm{N}}^{\mathrm{ESO}} < \mathrm{RT}$, there should be no magnetic ordering in the ESO. Furthermore, the paramagnetic (PM) behavior has also been observed from the linear behavior of the magnetization measurements at $\mathrm{RT}$.[19] Hence the stiffness constants of the ESO were determined in the paramagnetic phase (Co, Ni and Cu ions are randomly populated with positive and negative moments with net magnetic moment $= 0$) and the nonmagnetic (NM) phase (individual moments on Co, Ni and Co ions $= 0$). The stiffness constants of the ESO in the AFM ordered phase were also determined for reference. Similarly, in the case of CuO, $T_{\mathrm{N}}^{\mathrm{CuO}} < \mathrm{RT}$, we determined the elastic constants of CuO in the nonmagnetic cubic phase. CuO in the monoclinic phase with AFM ordering at low temperature were also determined for comparison, while the stiffness constants of NiO and CoO were determined in the AFM ordered cubic phase.

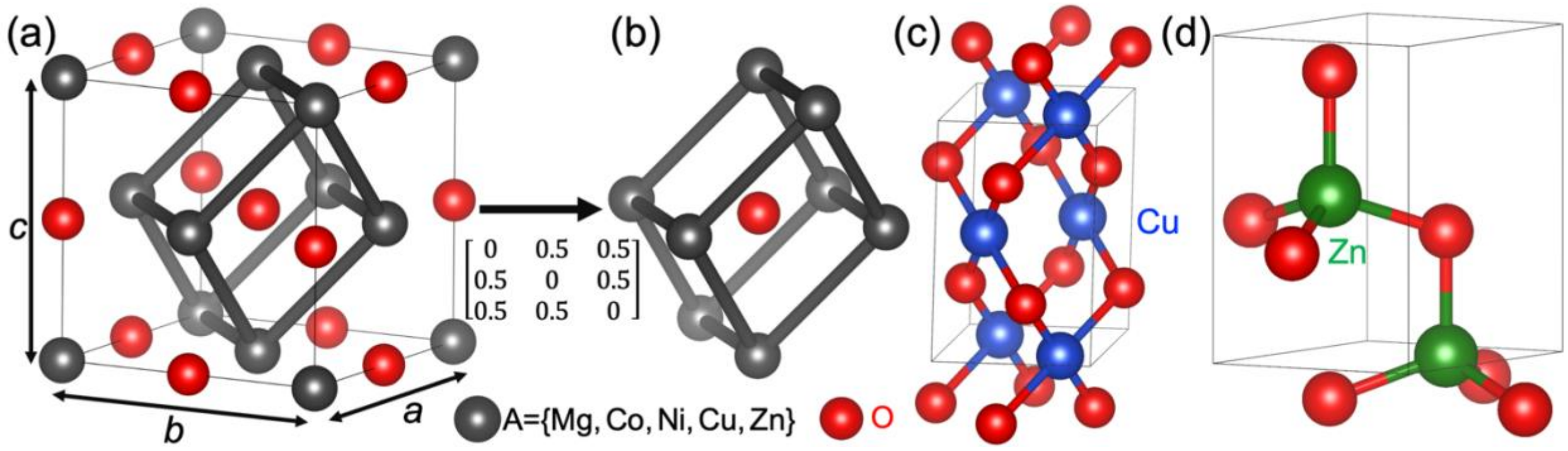

Figure 1: (a) Face centered cubic structure of rock salt phase with four lattice points of entropy stabilized oxide. (b) Primitive cell of rock salt structure with four lattice points obtained from rotation matrix shown near arrow. A-site is quasi-randomly occupied with equal probability by Mg, Co, Ni, Cu, Zn. A $4 \times 4 \times 5$ supercell of the primitive cell, containing 160 atoms was used to model the ESO. (c) tenorite and (d) wurtzite structures.

## III. RESULTS AND DISCUSSION

Figure 2 shows the measured indentation modulus as a function of $l_1^2 l_2^2 + l_1^2 l_3^2 + l_2^2 l_3^2$, where $l_i$ are the direction cosines. Each data point corresponds to the average of all indents obtained within a given grain. The error bars correspond to one standard deviation about the mean value. Two data sets are included in the plot for peak indentation loads of 2 and 5 mN. A regression analysis of the results indicates that the indentation modulus is effectively independent of crystallographic orientation. Also included in Figure 2 are the Young's modulus estimated from DFT calculations and estimated values for the indentation modulus from first principles according to the formulation of Vlassak and Nix for grains oriented in the [100], [110] and [111] directions. Details of the data analysis are included as supplemental material. The blue-solid, red-dotted and black-dashed lines corresponds to the DFT-estimated Young's modulus for antiferromagnetic, paramagnetic and nonmagnetic structures, respectively. The elastic constants for the paramagnetic structure were obtained from Boltzmann average of two different random structures. Refer to the Sec. II of supplemental material for detailed method of determining elastic constants and Young's modulus. The magnitude of the predicted Young's modulus was found to be greater than the experimentally-determined indentation modulus. It has been found that among the various techniques available for measuring Young's modulus, nanoindentation usually has the lowest precision[41] and that differences between measured and estimated values of Young's modulus can vary by as much as 20%. It is believed that the precision of the nanoidentation measurement contributed to these discrepancies. Furthermore, we find that the predicted elastic constants for the parent oxides were generally greater than those measured experimentally. Refer to Table SIV of supplemental material for a comprehensive list of experimentally determined elastic stiffness constants ($c_{ij}$), anisotropy factors ($A^{\mathrm{Z}}, A^{\mathrm{U}}$), Young's modulus ($E$), Poisson's ratio ($\nu$), bulk and shear modulus ($K, G$) of constituent binary oxides.[37,38,42–49]

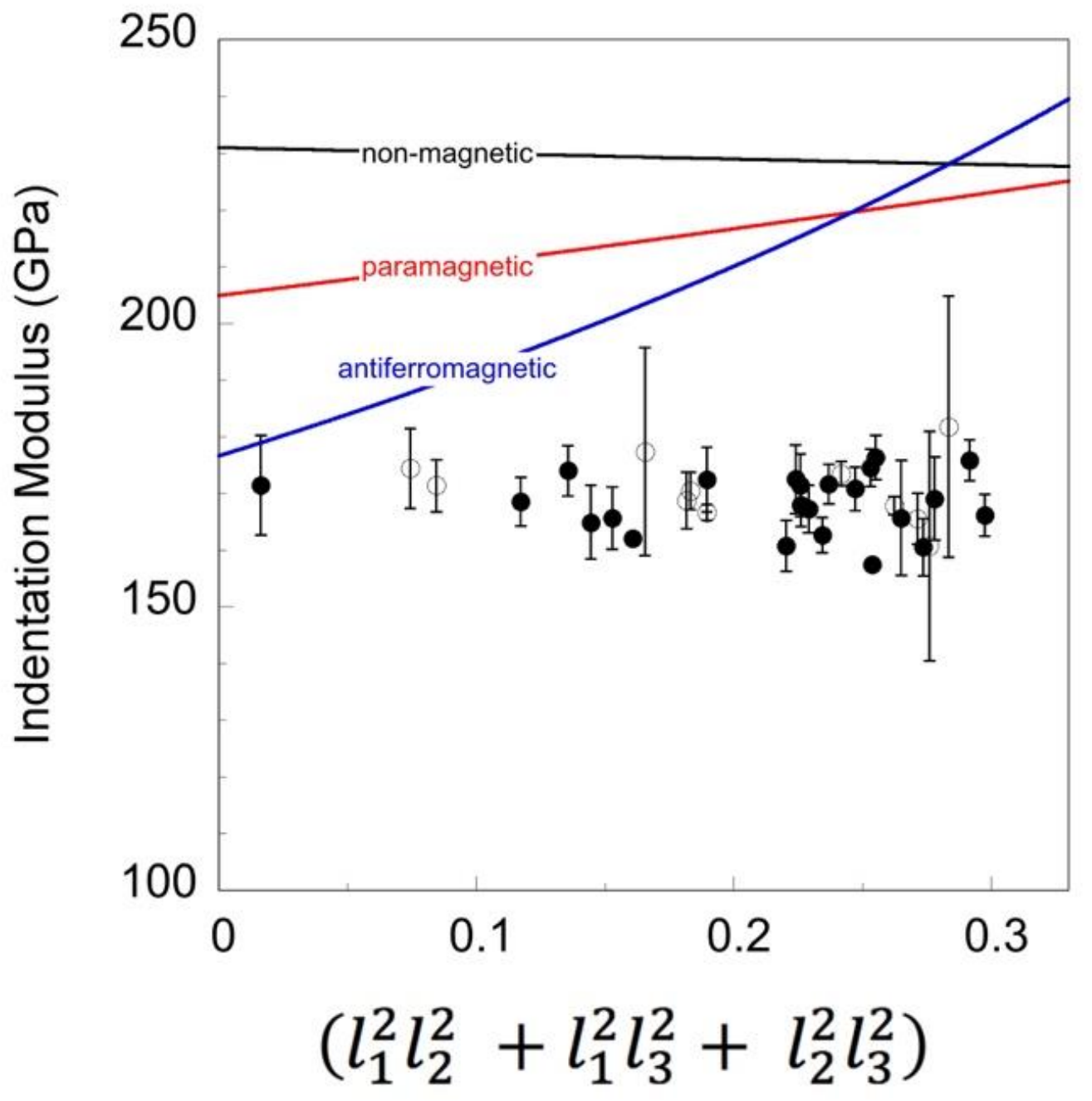

Figure 2. Indentation modulus of (Mg, Co, Ni, Cu, Zn)O entropy stabilized oxide as a function of crystallographic orientation. The two data sets (open and closed symbols) correspond to peak loads of 2 and 5 mN, respectively. Data points correspond to average values from all indents on an individual grain. The crystallographic orientation of each grain was determined using electron backscatter diffraction. The blue-solid, red-dotted and black-dashed lines correspond to the DFT-estimated Young's modulus for antiferromagnetic, paramagnetic and nonmagnetic structures, respectively.

To better understand the measured isotropic behavior we employed first principles calculations to predict the elastic constants for the binary oxides and the ESO. Table I lists calculated values for the stiffness constants, $c_{ij}$, bulk modulus, $K_{\mathrm{VRH}}$, shear modulus, $G_{\mathrm{VRH}}$, elastic modulus, $E$, Poisson's ratio, $\nu$, Zenner anisotropy ratio, $A^{\mathrm{Z}}$, and universal anisotropy ratio, $A^{\mathrm{U}}$. Refer to Section II and Table SII of the supplemental material for detailed discussion on the method of estimation of the elastic constants from the first principles. For consistency in the comparison of elastic properties between the constituent binary oxides and the ESO, calculations were performed for CuO and ZnO in the rock-salt structure. In all cases we studied, including the AFM, NM and PM structures the following relations hold:

$$c_{44} > 0$$

$$c_{11} > |c_{12}|$$

$$c_{11} + 2c_{12} > 0$$

which are conditions for mechanical stability for cubic crystals.[50]

First, we examine the Zener's anisotropy ratio, $A^{\mathrm{Z}}$, for the individual binary oxides. We find that $A^{\mathrm{Z}} < 1$ for AFM ordered CoO and NiO, and $A^{\mathrm{Z}} > 1$ for NM MgO, CuO and ZnO. Consequently, the AFM ordered materials – CoO and NiO – are stiffer towards shearing in the $(110)$ planes along the $[1,-1,0]$ direction than shearing in the $(100)$ plane along the $[0,1,0]$ direction. Furthermore, the opposite is true for the NM materials – MgO, CuO and ZnO – in the cubic phase, i.e., they are stiffer towards shearing in the $(100)$ planes along the $[0,1,0]$ direction than shearing in the $(110)$ plane along the $[1,-1,0]$ direction . The Young's modulus for individual binary oxides, along with the AFM, PM and NM models of ESO are plotted in Figure S3 of supplemental material.

Second, we constructed a disordered unit cell using the SQS approach, to study the elastic properties of the ESO. Here we considered a nonmagnetic phase and two different magnetic phases – AFM ordering similar to ground phases of CoO and NiO as well as a paramagnetic phase. We have considered two different SQS configurations for paramagnetic phase whose energy difference was within 20 meV. The relaxed structures of all entropy stabilized oxides can be found at the end of supplemental material. The value of $A^{\mathrm{Z}}$ for the ESO modelled using AFM, PM and NM phases are 1.42, 1.12 and 0.98, respectively. We find that the elastic anisotropy of the AFM phase is much stronger than those of the PM and NM phases. This suggests that the elastic anisotropy of the ESO is highly sensitive to its magnetic structure. Specifically, the elastic anisotropy of $\mathrm{AFM} > \mathrm{PM} > \mathrm{NM}$ phases and should generally decrease with increasing temperature, due to the AFM→PM phase transition. The influence of the magnetic state on the local structure and on the elastic properties is further analyzed through the radial distribution function, $g(r)$, and the polar displacements of the cations for each model.

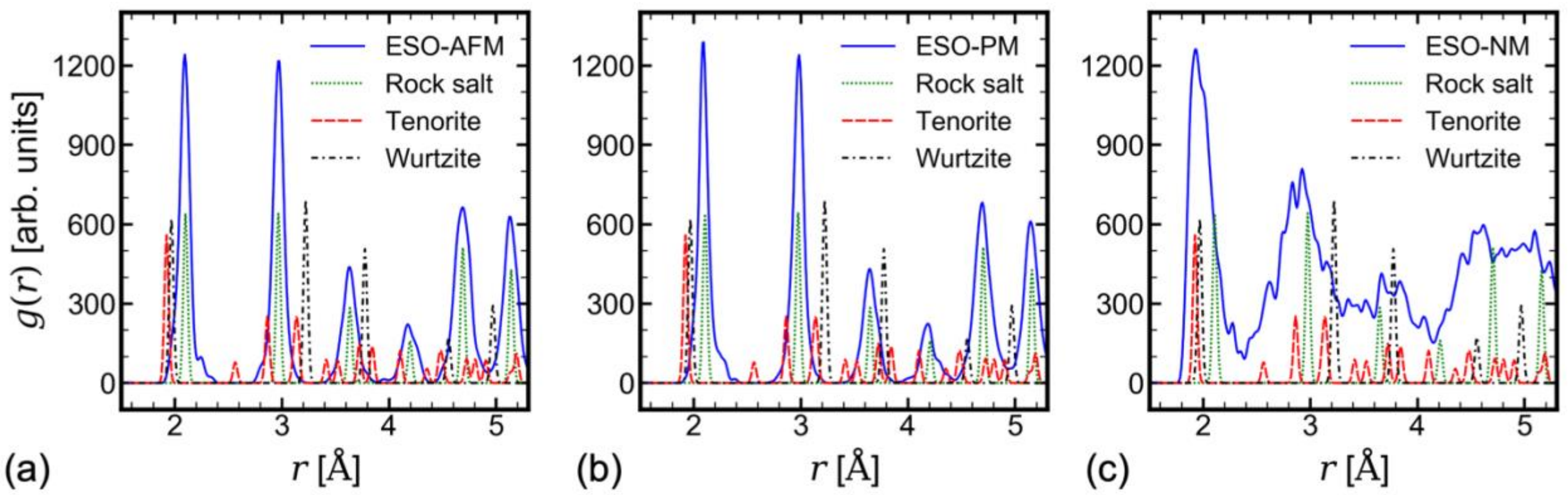

Figure 3. Radial distribution function, $g(r)$ is plotted in solid blue line for (a) AFM, (b) PM and (c) NM structures. For comparison $g(r)$ is also plotted for rock salt, tenorite and wurtzite structures with equivalent lattice constants, in dotted green line, dashed red line, and dot-dashed line, respectively.

Figure 3 shows $g(r)$ for (a) the AFM, (b) the lowest energy PM (of the two studied models) and (c) the NM structures, compared to representative $g(r)$ of rock salt at ESO lattice constant, tenorite at CuO lattice constant and wurtzite at Zn lattice constant. While there are only minor differences in $g(r)$ between the AFM and PM structures, we find significant differences of these structures with the NM structure. Specifically, the first blue peak at $\sim 2$ Å – corresponding to the $A - \mathrm{O}$ bond distance – aligns with the (dotted green) peak of the rock salt phase for the AFM and PM structures, suggesting pure rock salt character. On the contrary, the first blue peak of the NM structure is broadened with character comprising of rock salt, tenorite and wurtzite structures. This suggests that the presence of local distortions, arising from the tendency of Cu and Zn to form secondary phases, may enhance the isotropy of the material. Furthermore, the similarity between the AFM and PM structures suggests that the decrease in the value of $A^{\mathrm{Z}}$ between the AFM to PM structures, may exclusively be attributed to the strongly anisotropic magnetic interactions in the AFM model. We find similar trends for comparisons of the cation polar displacements (CPD) between the AFM, PM and NM structures. Figure S4 in supplemental material shows the distribution of CPD for all structures. The cation displacements in the AFM and PM structure are small for all cations $< 0.05$ Å. However, the cation displacements in the NM structure are substantially large ~$0.5$ Å for Cu and Zn cations, suggesting a tendency to locally distort to the tenorite and wurtzite structures, respectively. We also find that while Mg displacements are comparatively smaller, ~$0.25$ Å, Ni and Co are prone to local displacements.

Figure 4 shows a plot of the average bulk modulus versus the average shear modulus for the 5 constituent binary oxides and the ESO, (Mg, Co, Ni, Cu, Zn)O. Values for the elastic moduli of NiO and CoO were calculated considering AFM ordering and with $U = 6.0$ eV. The red line in the diagram highlights the boundary established by the largest and smallest average values of the bulk and shear moduli of the constituent binary oxides. It was found that the values predicted for the average moduli of the ESO, in the NM, AFM and PM magnetic states, falls within these bounds and that these values are comparable to the values calculated using the rule of mixtures (ROM), which is often used as a first approximation to estimate properties, such as the elastic properties, of random solid solutions[51]. While the structure of high-entropy metallic alloys is different from that of the ESO compound investigated in this paper, it is worth mentioning that many high-entropy alloys have been found to follow this trend,[51,52] although there are exceptions such as NiFeCrCoMn, which is predicted to have a bulk modulus lower than those of its constituents[52]. While alloying effects are known to result in alloys with elastic properties that are lower than those of their constituents, these trends have not been established for either high-entropy alloys or entropy-stabilized oxides.

Also included in Figure 4 are lines corresponding to different values of Pugh's ratio (K/G), including one for K/G=1.75, which is widely used to distinguish materials with ductile behavior (K/G>1.75). The results in Figure 4 suggests that the ESO compound, which has a value of Pugh's ratio – equal to 2.07, 2.20 and 2.30, in the NM, PM and AFM phases, respectively, should exhibit ductile behavior, which is consistent with the indentation behavior in that no cracks formed at the corners of the indents.

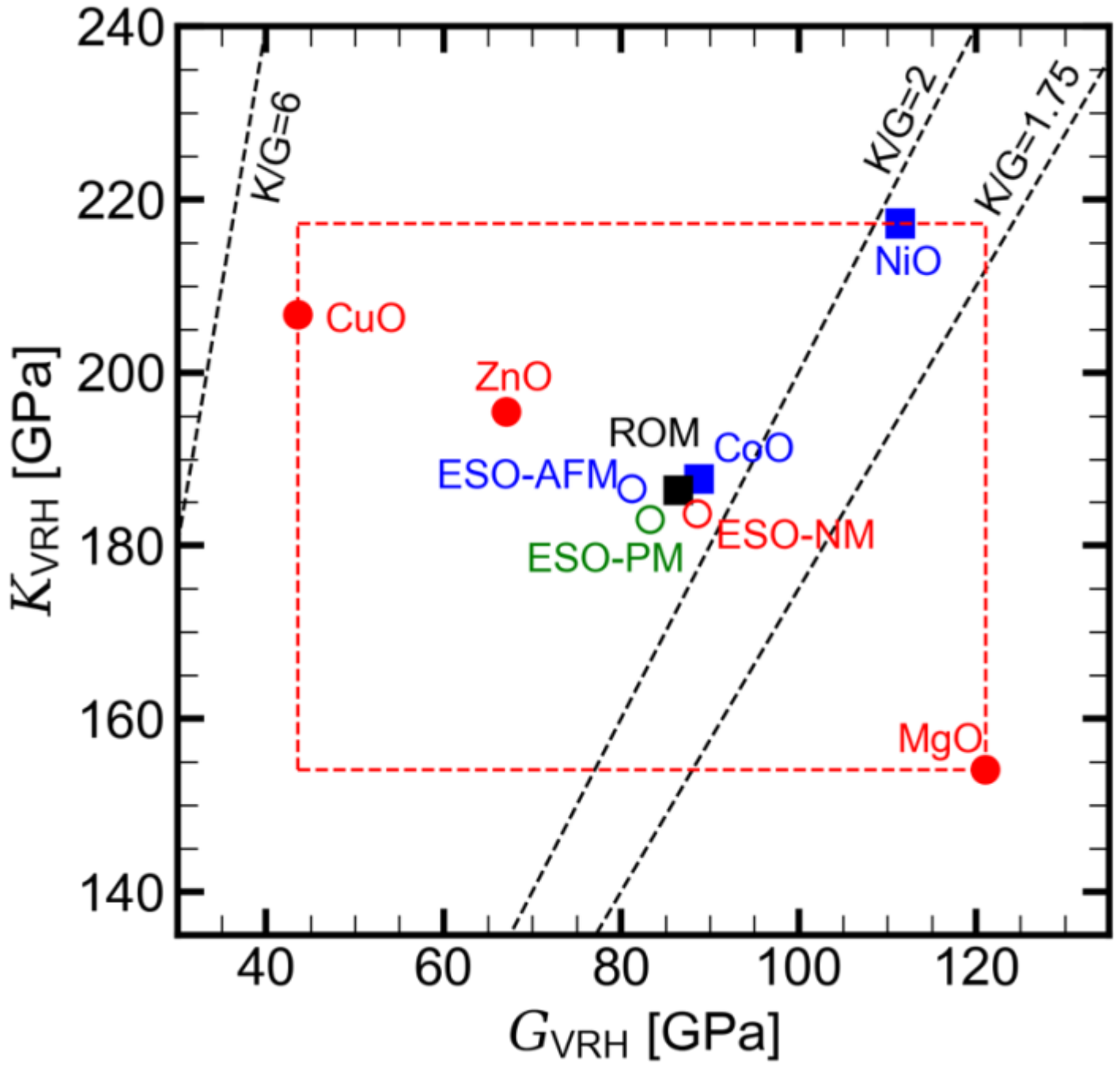

Figure 4. Plot of Voigt-Reuss-Hill Average values of bulk vs shear moduli. Data points correspond to values determined from first principles calculations. ROM value shown in black solid square corresponds to values calculated using the rule of mixtures. Red solid circles correspond to values calculated for non-magnetic binary oxides with $U = 0.0$ eV. Blue solid squares correspond to values calculated for AFM binary oxides with $U = 6.0$ eV. Values for the ESO compound are shown with open circles: (i) the red circle for nonmagnetic calculation at $U = 0.0$ eV, (ii) the green circle is for paramagnetic calculation with $U = 6.0$ eV, and (iii) the blue circle is for antiferromagnetic calculation with $U = 6.0$ eV. The dashed red line is used to illustrate bounds provided by the elastic constants of the constituent binary oxides.

Figure 5 plots the values of $A^{\mathrm{Z}}$ vs. $A^{\mathrm{U}}$ for the ESO in the AFM, PM and NM structures. The data for the constituent parent oxides are also shown for comparison. The line $A^{\mathrm{Z}} = 1$ corresponds to elastic isotropy and separates the plot into two regions: above this line $c_{44} > \frac{c_{11}-c_{12}}{2}$ while the opposite case holds below the line. It is worth noting that Zener's anisotropy parameter for the ESO is close to the boundary of the values of the constituent binary oxides and that an average of these values yields 1.04 (represented in the plot as ROM), which is very close to the value determined from first principle calculations (0.98). The calculated value of the elastic anisotropy index, $A^{\mathrm{U}}$ for the ESO was found to be effectively zero ($3.4 \times 10^{-4}$), which is representative of a material with high elastic isotropy. Conversely, for the ESO in the paramagnetic phase (which is stable at $T = \mathrm{RT}$) $A^{\mathrm{Z}} = 1.12$ indicating a mild degree of elastic anisotropy, which further increases to $A^{\mathrm{Z}} = 1.42$ for low temperature AFM phase, indicating enhanced anisotropy.

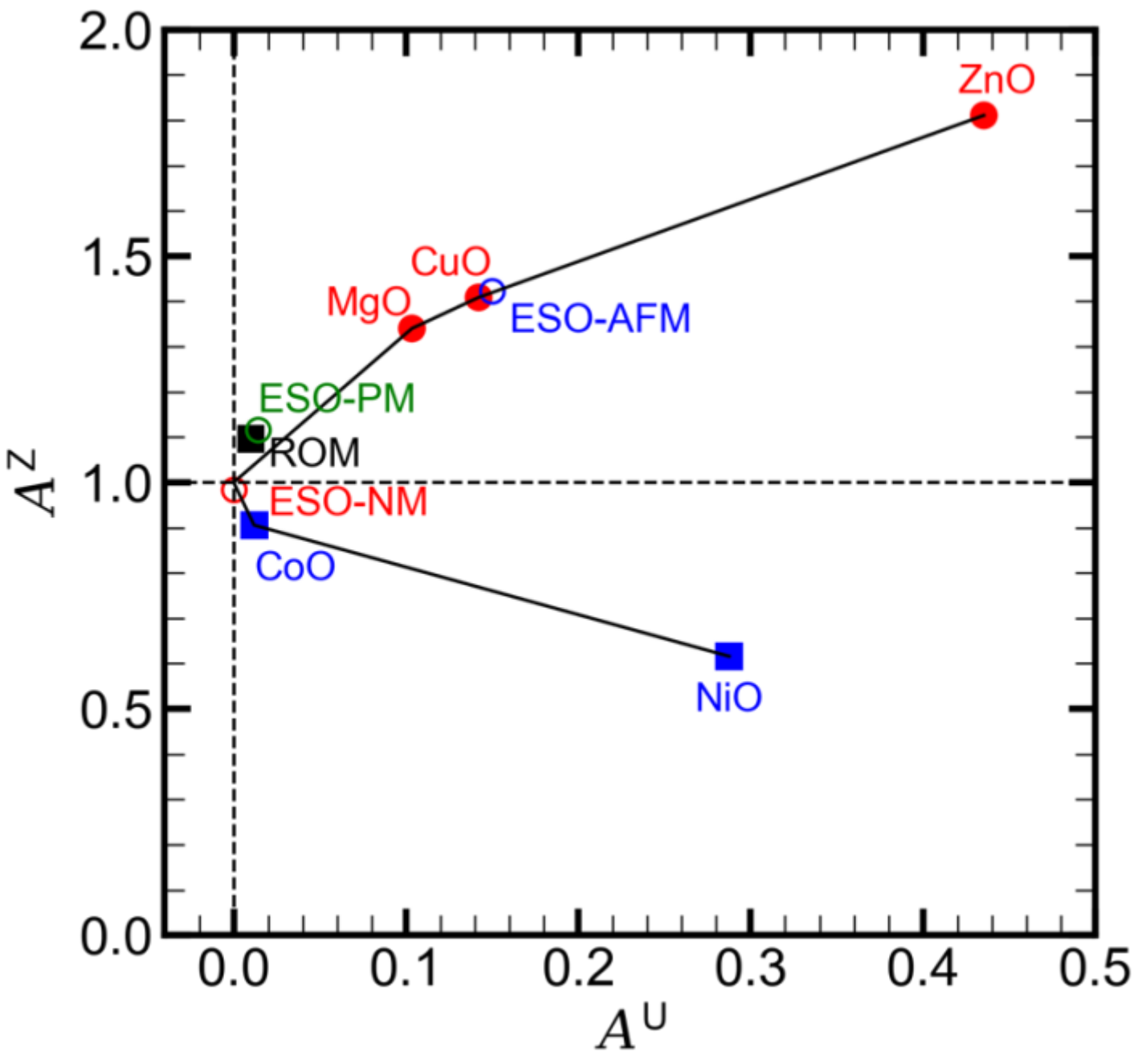

Figure 5. Plot of Zener elastic anisotropy parameter $A^{\mathrm{Z}}$ vs. elastic anisotropy index $A^{\mathrm{U}}$, for the ESO and its constituent binary oxides. Solid red circles correspond to values determined using $U = 0.0$ eV for nonmagnetic calculations. Solid blue squares correspond to calculations for CoO and NiO for $U = 6.0$ eV and considering AFM ordering. Open symbols correspond to values for the ESO compound: (i) the red circle is for NM calculation with U=0.0 eV, (ii) the green circle is for PM calculation with $U = 6.0$ eV and (iii) the blue circle is for AFM calculation with $U = 6.0$ eV. The black solid lines are used to identify the boundary set by the constituent binary oxides. The elastic anisotropy values of PM and NM calculations for the ESO compound are approximately equal to the average values (shown in solid black square) obtained from the rule-of mixtures.

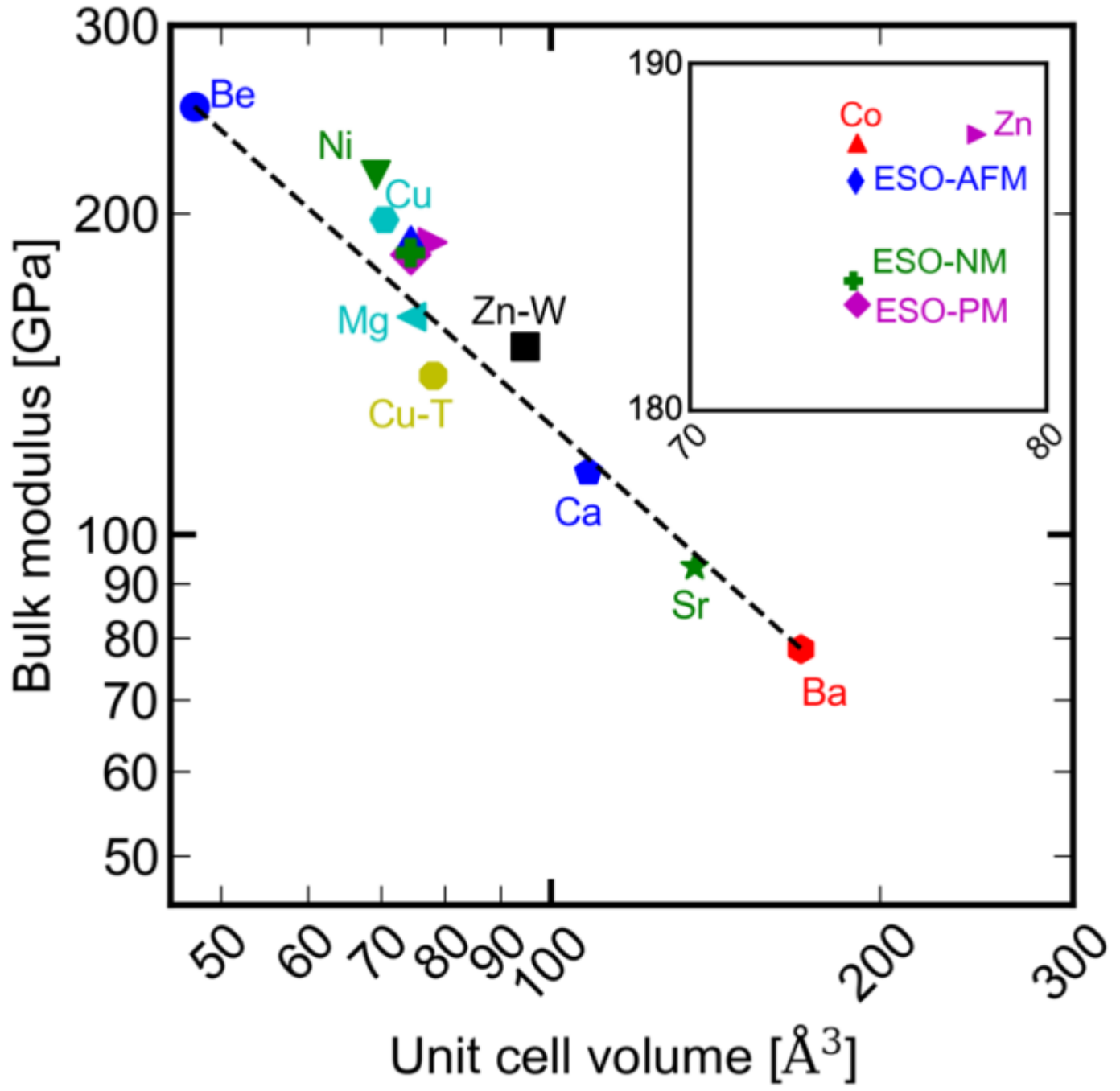

Figure 6: Bulk modulus plotted as a function of unit cell volume plotted on a log-log scale for several divalent binary oxides in rock salt phase including ESO-AFM, ESO-PM and ESO-NM. CuO and ZnO are plotted in their respective ground phases represented by Cu-T (tenorite) and Zn-W (wurtzite). Inset shows the data points clustered near CuO.

Finally, we analyze the bulk modulus of the ESO through the well-known relation between the unit cell volume and bulk modulus in ionic solids.[53,54] Figure 6 plots the VRH-averaged bulk modulus as a function of the unit cell volume, for several divalent binary oxides in the rock salt phase, including ESO-AFM, ESO-PM and ESO-NM. We find that the bulk constants of ESO in the AFM, PM and NM structures, follow the general trend dictated by the binary oxides of the divalent cations.

## Conclusions

The elastic constants of the multi-component oxide (Mg, Co, Ni, Cu, Zn)O were determined form first principles calculations and experimental indentation measurements. The values of indentation modulus at ambient conditions revealed a high degree of elastic isotropy. Predictions using first principles calculations confirmed a high degree of isotropy for a nonmagnetic structure with $A^{\mathrm{Z}} = 0.98$ and slightly increases for a paramagnetic structure $A^{\mathrm{Z}} = 1.12$. Conversely, the low temperature antiferromagnetic phase, has a higher degree of anisotropy, $A^{\mathrm{Z}} = 1.42$. Comparisons of the AFM, PM and NM structures indicate two complementary contributions, influencing the isotropy. First, the uncorrelated magnetic state promote isotropy, which can be further amplified by inclusion of local wurtzite and tenorite structures. Interestingly, we find that the bulk and shear modulus of the multicomponent oxide are within the boundaries set by the constituent oxides and close to the values predicted by the rule of mixtures. Computational data and experimental observations indicate that the ESO should exhibit ductile behavior, which would make it an attractive candidate for engineering applications.

## ACKNOWLEDGEMENTS

Research sponsored by the Laboratory Directed Research and Development Program of Oak Ridge National Laboratory, managed by UT-Battelle, LLC, for the U. S. Department of Energy. The authors acknowledge the computational resources at NERSC and their colleagues Christopher Fancher and Raphael P. Hermann of ORNL for reviewing the manuscript. De Xin Chen participated in this project through a Mickey Leland Energy Fellowship (MLEF) internship at the Oak Ridge National Laboratory while he was an undergraduate student at the Massachusetts Institute of Technology. The MLEF program is funded by the Fossil Energy Program at the U.S. Department of Energy Program and managed by Oak Ridge Associated Universities.

Table I: Values of elastic constants, bulk modulus, shear modulus, Young's modulus, Poisson's ratio, Zener anisotropy ratio and universal anisotropy ratio for individual binary oxides and ESO. Reference [0] corresponds to this work.

| Compound | Phase | Reference | Method | $c_{11}$ / $c_{22}$ / $c_{33}$ | $c_{12}$ / $c_{23}$ / $c_{31}$ | $c_{44}$ / $c_{55}$ / $c_{66}$ | $K_{\mathrm{VRH}}$ | $G_{\mathrm{VRH}}$ | $E$ | $\nu$ | $A^{\mathrm{Z}}$ | $A^{\mathrm{U}}$ |
|---|---|---|---|---|---|---|---|---|---|---|---|---|
| | | | | [GPa] | [GPa] | [GPa] | [GPa] | [GPa] | [GPa] | - | - | - |
| MgO | Cubic-NM | [0] | DFT; U=0 | 289.53 | 86.37 | 136.10 | 154.09 | 121.05 | 287.80 | 0.19 | 1.34 | 0.10 |
| CoO | Cubic-NM | [0] | DFT; U=0 | 395.00 | 191.00 | 36.00 | 259.00 | 55.49 | 155.36 | 0.40 | 0.35 | 1.42 |
| CoO | Cubic-AFM | [0] | DFT; U=0 | 313.37 | 146.36 | 59.81 | 202.03 | 68.38 | 184.34 | 0.35 | 0.72 | 0.14 |
| CoO | Cubic-AFM | [0] | DFT; U=6 | 313.36 | 124.81 | 85.36 | 187.66 | 88.82 | 230.15 | 0.30 | 0.91 | 0.01 |
| NiO | Cubic-NM | [0] | DFT; U=0 | 360.00 | 205.00 | 36.00 | 256.67 | 49.21 | 138.75 | 0.41 | 0.47 | 0.74 |
| NiO | Cubic-AFM | [0] | DFT; U=0 | 389.16 | 131.56 | 43.09 | 217.47 | 68.05 | 184.86 | 0.36 | 0.34 | 1.59 |
| NiO | Cubic-AFM | [0] | DFT; U=6 | 415.72 | 117.97 | 91.69 | 217.22 | 111.45 | 285.51 | 0.28 | 0.62 | 0.29 |
| CuO | Cubic-NM | [0] | DFT; U=0 | 254.00 | 183.00 | 50.00 | 206.67 | 43.59 | 122.18 | 0.40 | 1.41 | 0.14 |
| CuO | Cc-AFM | [0] | DFT; U=0 | 154.44 / 151.31 / 367.83 | 144.77 / 143.84 / 143.60 | 67.82 / 8.13 / 9.91 | 159.73 | 21.62 | 62.07 | 0.44 | - | 11.81 |
| CuO | Cc-AFM | [0] | DFT; U=6 | 295.68 / 126.74 / 327.79 | 152.58 / 85.59 / 151.57 | 100.65 / 18.35 / 56.60 | 140.87 | 46.77 | 126.32 | 0.35 | - | 4.12 |
| ZnO | Cubic-NM | [0] | DFT | 258.13 | 164.10 | 85.12 | 195.45 | 67.08 | 180.58 | 0.35 | 1.81 | 0.44 |
| ZnO | Hexagonal -NM | [0] | DFT | 204.89 / 204.89 / 220.11 | 110.98 / 125.58 / 125.58 | 39.65 / 36.98 / 36.98 | 150.07 | 40.39 | 111.19 | 0.38 | - | 0.05 |
| ESO | PM [160] | [0] | DFT; U=6 | 287.00 | 131.00 | 87.00 | 183.00 | 83.28 | 216.94 | 0.30 | 1.12 | 0.01 |
| ESO | NM [160] | [0] | DFT; U=0 | 303.00 | 124.00 | 88.00 | 183.67 | 88.60 | 228.97 | 0.29 | 0.98 | 0.00 |
| ESO | AFM [160] | [0] | DFT; U=6 | 274.27 | 142.70 | 93.53 | 186.56 | 81.23 | 212.80 | 0.31 | 1.42 | 0.15 |

**FIGURE CAPTIONS**

Figure 2: (a) Face centered cubic structure of rock salt phase with four lattice points of entropy stabilized oxide. (b) Primitive cell of rock salt structure with four lattice points obtained from rotation matrix shown near arrow. A-site is quasi-randomly occupied with equal probability by Mg, Co, Ni, Cu, Zn. A $4 \times 4 \times 5$ supercell of the primitive cell, containing 160 atoms was used to model the ESO. (c) tenorite and (d) wurtzite structures.

Figure 2. Indentation modulus of (Mg, Co, Ni, Cu, Zn)O entropy stabilized oxide as a function of crystallographic orientation. The two data sets correspond to peak loads of 2 and 5 mN. Data points correspond to average values from measurements on individual grains. The crystallographic orientation of each grain was determined using electron backscatter diffraction. The blue-solid, red-dotted and black-dashed lines corresponds to the DFT-estimated Young's modulus for antiferromagnetic, paramagnetic and nonmagnetic structures, respectively.

Figure 3. Radial distribution function, $g(r)$ is plotted in solid blue line for (a) AFM, (b) PM and (c) NM structures. For comparison $g(r)$ is also plotted for rock salt, tenorite and wurtzite structures with equivalent lattice constants, in dotted green line, dashed red line, and dot-dashed line, respectively.

Figure 4. Plot of Voigt-Reuss-Hill Average values of bulk vs shear moduli. Data points correspond to values determined from first principles calculations. ROM value shown in black solid square corresponds to values calculated using the rule of mixtures. Red solid circles correspond to values calculated for non-magnetic binary oxides with $U = 0.0$ eV. Blue solid squares correspond to values calculated for AFM binary oxides with $U = 6.0$ eV. Values for the ESO compound are shown with open circles: (i) the red circle for nonmagnetic calculation at $U = 0.0$ eV, (ii) the green circle is for paramagnetic calculation with $U = 6.0$ eV, and (iii) the blue circle is for antiferromagnetic calculation with $U = 6.0$ eV. The dashed red line is used to illustrate bounds provided by the elastic constants of the constituent binary oxides.

Figure 5. Plot of Zener elastic anisotropy parameter $A^{\mathrm{Z}}$ vs. elastic anisotropy index $A^{\mathrm{U}}$, for the ESO and its constituent binary oxides. Solid red circles correspond to values determined using $U = 0.0$ eV for nonmagnetic calculations. Solid blue squares correspond to calculations for CoO and NiO for $U = 6.0$ eV and considering AFM ordering. Open symbols correspond to values for the ESO compound: (i) the red circle is for NM calculation with U=0.0 eV, (ii) the green circle is for PM calculation with $U = 6.0$ eV and (iii) the blue circle is for AFM calculation with $U = 6.0$ eV. The black solid lines are used to highlight the boundary set by the constituent binary oxides. The elastic anisotropy values of PM and NM calculations for the ESO compound are approximately equal to the average values (shown in solid black square) obtained from the rule-of mixtures.

Figure 6: Bulk modulus plotted as a function of unit cell volume plotted on a log-log scale for several divalent binary oxides in rock salt phase including ESO-AFM, ESO-PM and ESO-NM. CuO and ZnO are plotted in their respective ground phases represented by Cu-T (tenorite) and Zn-W (wurtzite). Inset shows the data points clustered near CuO.

**SUPPLEMENTAL MATERIAL**

## I. Characterization

A piece from the sintered ESO sample was prepared metallographically by a sequence of grinding and polishing steps using increasingly finer polishing media. The final preparation step consisted of polishing using submicron colloidal silica. The microstructure and crystallographic composition of the material was characterized using a Field Emission Scanning Electron Microscope (JEOL JSM-6500F equipped with a Hikari Pro detector) at an operating current of 4μA and a voltage of 20kV. Figure S1. Shows an electron backscatter diffraction micrograph of ESO sample showing array of nanoindents across several grains. The color of each grain corresponds to a different crystallographic orientation.

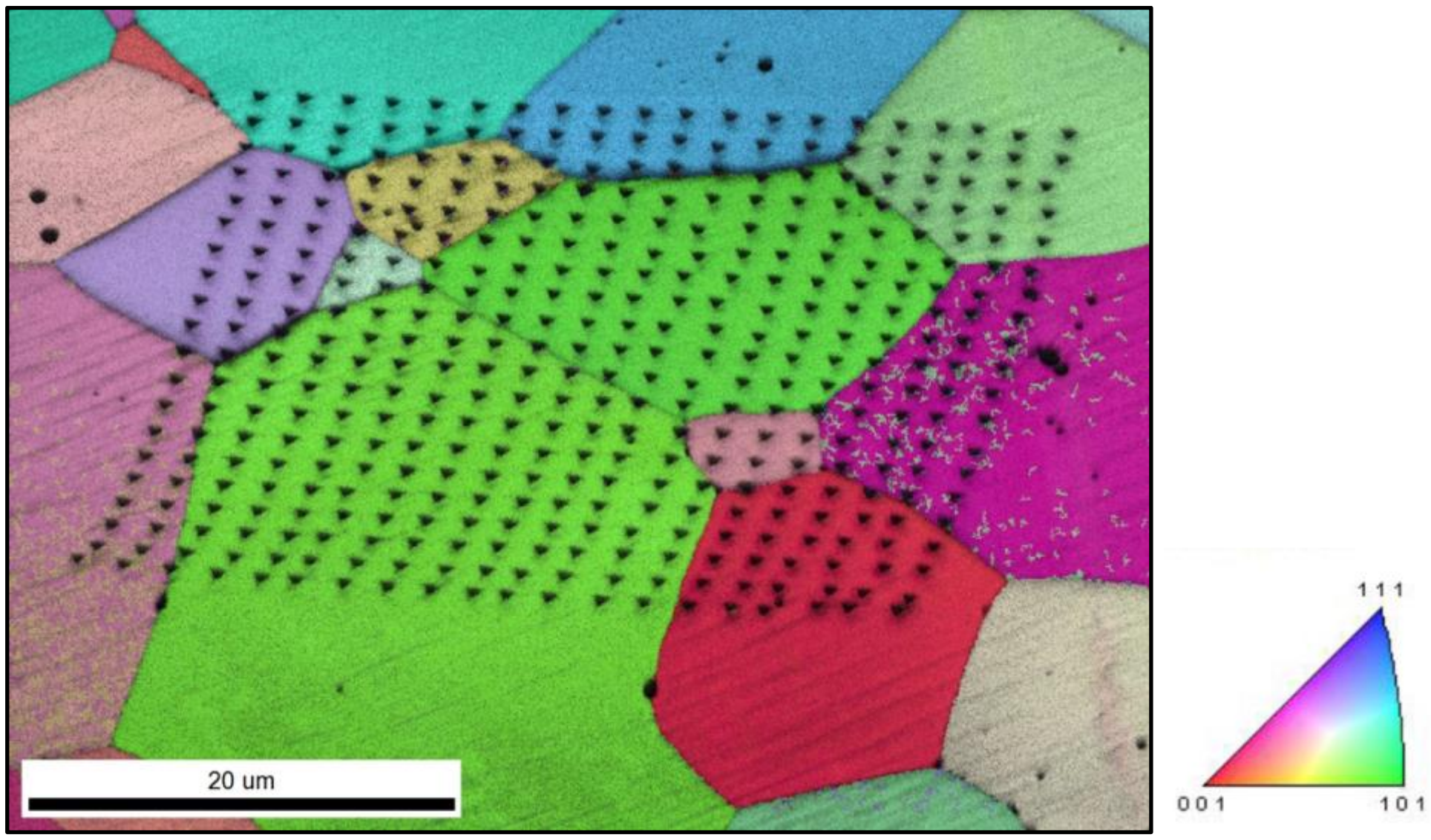

Figure S1. Electron backscatter diffraction micrograph of ESO sample showing array of nanoindents across several grains. The color of each grain corresponds to a different crystallographic orientation.

## II. Indentation

The indents, which were spaced apart from each other to avoid interaction of their deformation fields, were obtained using a nanoindenter (TI 950 TriboIndenter, Bruker, Minneapolis, MN 55344). A trapezoidal profile with loading and unloading segments lasting 5 seconds and a 2-second hold segment was used. Experiments were performed at peak loads of both 2mN and 5 mN. Figure S1 shows an electron backscatter diffraction micrograph of nanoindents obtained on the surface of several grains each with a different crystallographic orientation. Several arrays of indents were obtained to ensure good coverage of crystallographic orientations between the (001) and (111) directions.

Figure S2 illustrates the procedure followed to compare experimentally-determined values of the indentation modulus with those predicted from first principles.

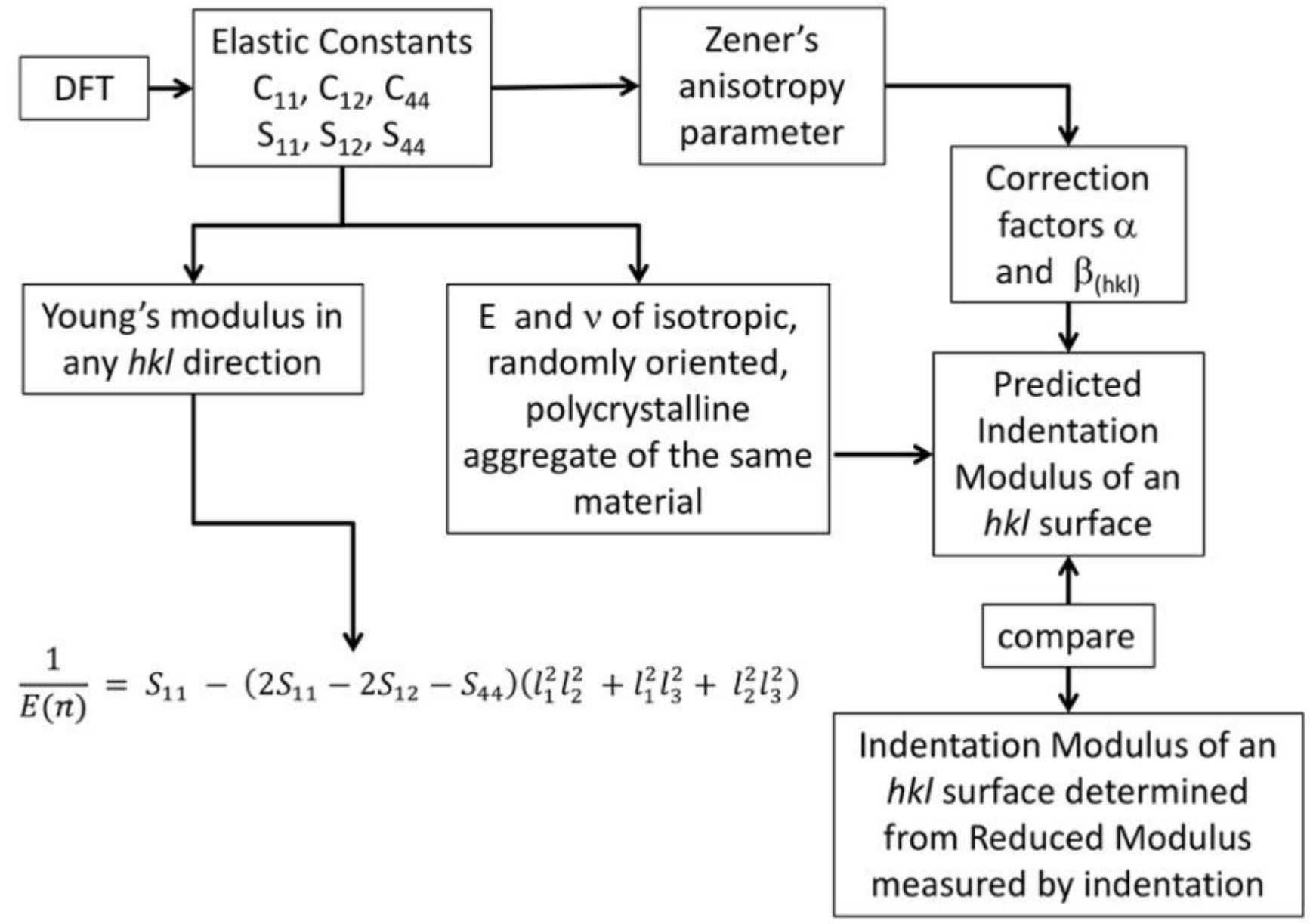

Figure S2. Schematic showing the procedure for comparing experimentally-determined values of the indentation modulus with those predicted from first principles.

The reduced modulus of elastically isotropic materials is given by

$$\frac{1}{E_r} = \left(\frac{1-\nu^2}{E}\right) + \left(\frac{1-\nu_i^2}{E_i}\right) \quad \text{(S1)}$$

where $E$ and $\nu$ are Young's modulus and Poisson's ratio for the specimen, and $E_i$ and $\nu_i$ are the equivalent quantities for the diamond indenter (for diamond, $E_i = 1141$ GPa and $\nu_i = 0.07$).

For anisotropic materials [1], Equation (S1) is transformed into:

$$\frac{1}{M_r} = \left(\frac{1}{M}\right) + \left(\frac{1-\nu_i^2}{E_i}\right) \tag{S2}$$

According to the analysis of Vlassak and Nix [1], the indentation modulus on a surface with crystallographic orientation $\{hkl\}$ is given by

$$M_{hkl} = \alpha\, \beta_{hkl}\, \left(\frac{E}{1-\nu^2}\right)_{isotropic} \tag{S3}$$

where $\alpha$ is a correction factor to account for the geometry of the indenter tip and its orientation with respect to the crystallographic orientation of the indented surface. For triangular indenters $\alpha$ ~ 1.058 [1]. The correction factor $\beta_{hkl}$ for crystals with cubic symmetry is given by:

$$\beta_{hkl} = a + c(A^{Z} - A_0)^B \tag{S4}$$

where $a$, $c$, $A_0$ and $B$ depend only on Poisson's ratio in the cube's directions, which in this case is given by [2].

$$\nu_{<100>} = -\frac{s_{12}}{s_{11}}$$

where $s_{ij}$ are compliance constants determined from first principles. Using values calculated for the compliance matrix of the non-magnetic phase of the EOS, which are listed in Table I of the main manuscript, we find $\nu_{\langle 100\rangle} = 0.313$. Values of $a$, $c$, $A_0$ and $B$, and therefore the correction factor $\beta_{(hkl)}$, which are listed in Table SI, were determined from Table I in Ref. [1], by linearly interpolating between $\nu_{\langle 100\rangle} = 0.3$ and $\nu_{\langle 100\rangle} = 0.35$. Values of Young's modulus and Poisson's ratio of an isotropic, randomly oriented, polycrystalline aggregate of the non-magnetic phase of the EOS can be calculated according to expressions listed in Table SIV to obtain:

$E_{isotropic}$ = 228.9 GPa

$\nu_{isotropic}$ = 0.292

Using Equation (S3) we obtain values for $M_{hkl}$, which are listed in Table SI and included in Figure 2 of the manuscript

Table SI. Correction factors to determine predicted value of indentation modulus

| $(hkl)$ | $a$ | $c$ | $A_0$ | $B$ | $\beta_{(hkl)}$ | $M_{(hkl)}$ |
|---|---|---|---|---|---|---|
| (1 0 0) | 1.266 | -0.265 | 0.000 | 0.288 | 1.002 | 264.8 GPa |
| (1 1 0) | 0.730 | 0.233 | -1.572 | 0.156 | 1.000 | 264.8 GPa |
| (1 1 1) | 0.744 | 0.254 | -0.021 | 0.188 | 0.998 | 264.8 GPa |

III. **Method for Determining Elastic Constants**

The linear elastic stiffness coefficients were calculated from DFT using the constitutive stress-strain relation for a crystal, irrespective of its symmetry. This method is similar to the one used in calculating the elastic constants presented in Materials Project (MP).[3,4] The following equation shows the constitutive stress-strain relation within the linear elastic limit.[5]

$$\sigma_i = c_{ij}\varepsilon_j \tag{S5}$$

$\sigma_i$, $\varepsilon_j$ and $c_{ij}$ are the stress, strain and elastic stiffness tensors in reduced Voigt notation,[5] respectively. The indices $i, j$ take the values between 1 and 6. Conversely, the elastic compliance tensor, $s_{ij}$, relates stress and strain components through $\varepsilon_i = s_{ij}\sigma_j$. Furthermore, the stiffness and compliance tensors are inversely related, $s_{ij} = \left[c_{ij}\right]^{-1}$.

For each oxide structure, we start from the relaxed lattice constants and generate a set of distorted structures by applying one of the six strain components, $\varepsilon_i$ at different magnitudes. For example, in the case of cubic structure, by choosing a positive (negative) value for $\varepsilon_1$ strain component – while fixing all other components to zero – the structure is expanded (compressed) only in the [1,0,0] direction. For each distorted structure representing a specific strain state $\varepsilon_i$, the six stress components in Voigt notation are obtained from a DFT calculation, in which, the lattice constants are fixed, and the ionic positions are relaxed. For each individually applied strain state, for example $\varepsilon_1$, six linear equations, $\sigma_i = c_{i1}\varepsilon_1$ are obtained from $\sigma_i = c_{ij}\varepsilon_j$. Consequently, a column of $c_{ij}$ constants are obtained for each strain component by fitting the stress–strain data to $\sigma_i = c_{ij}\varepsilon_j$. Table SII presents relevant elastic properties along with their descriptions and the equations used to derive the values from stiffness tensor $c_{ij}$. Specifically, elastic compliance constants $s_{ij}$, Voigt average of bulk and shear modulus, $K_{\mathrm{V}}, G_{\mathrm{V}}$, Reuss average of bulk and shear modulus, $K_{\mathrm{R}}, G_{\mathrm{R}}$, Voigt-Reuss-Hill average of bulk and shear modulus $K_{\mathrm{VRH}}, G_{\mathrm{VRH}}$, isotropic Young's modulus $E$, isotropic Poisson's ratio $\nu$, universal anisotropic ratio, $A^{\mathrm{U}}$, cubic anisotropic ratio, $A^{\mathrm{Z}}$, and directional dependence of Young's modulus as a function of direction cosines $E(l_1, m_1, n_1)$.

In the case of the ESO, even though the high temperature phase is experimentally found to be a cubic phase, in our 160-atoms model, the symmetry of the model is completely broken by the disorder on the cation sublattice. Nevertheless, the relaxed lattice constants of the ESO model were found to have small deviations from the perfect cubic lattice constants, where $a = b = c$ and $\alpha = \beta = \gamma = 90°$. Specifically, the standard deviation in $a, b, c$ is $< 5\%$ and $\alpha = \beta = \gamma$ is $<$ $1°$. Hence for the ESO models considered, the elastic constants were calculated by approximating the lattice to a cubic lattice. Furthermore, the elastic stiffness constant tensor of cubic phase materials, has only three independent values, $c_{11}$, $c_{12}$, $c_{44}$. The corresponding constitutive stress-strain relation for a cubic phase material, is shown in Eq.

$$\begin{bmatrix} \sigma_1 \\ \sigma_2 \\ \sigma_3 \\ \sigma_4 \\ \sigma_5 \\ \sigma_6 \end{bmatrix} = \begin{bmatrix} c_{11} & c_{12} & c_{12} & 0 & 0 & 0 \\ c_{12} & c_{11} & c_{12} & 0 & 0 & 0 \\ c_{12} & c_{12} & c_{11} & 0 & 0 & 0 \\ 0 & 0 & 0 & c_{44} & 0 & 0 \\ 0 & 0 & 0 & 0 & c_{44} & 0 \\ 0 & 0 & 0 & 0 & 0 & c_{44} \end{bmatrix} \begin{bmatrix} \varepsilon_1 \\ \varepsilon_2 \\ \varepsilon_3 \\ \varepsilon_4 \\ \varepsilon_5 \\ \varepsilon_6 \end{bmatrix} \tag{S6}$$

The Young's modulus as a function of parameter $(l_1^2 l_2^2 + l_1^2 l_3^2 + l_2^2 l_3^2)$, where $l_i$ are the directional cosines, can be estimated by:

$$E(l_1^2 l_2^2 + l_1^2 l_3^2 + l_2^2 l_3^2) = \left[s_{11} - 2\left(s_{11} - s_{12} - \tfrac{1}{2} s_{44}\right)(l_1^2 l_2^2 + l_2^2 l_3^2 + l_3^2 l_1^2)\right]^{-1} \tag{S7}$$

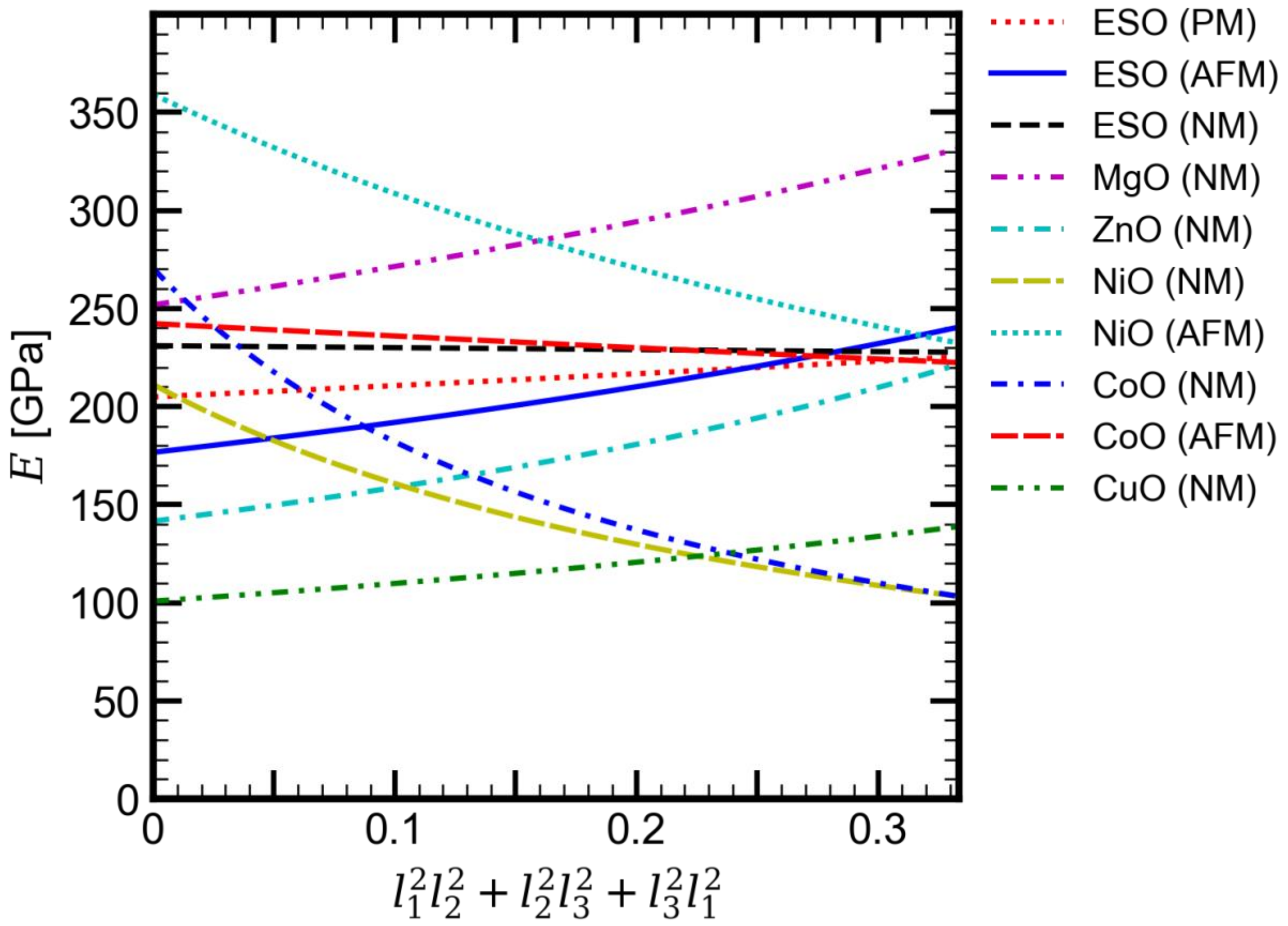

Figure S3. The Young's modulus as a function of $l_1^2 l_2^2 + l_2^2 l_3^2 + l_3^2 l_1^2$ for individual binary oxides, along with the AFM, PM and NM models of ESO.

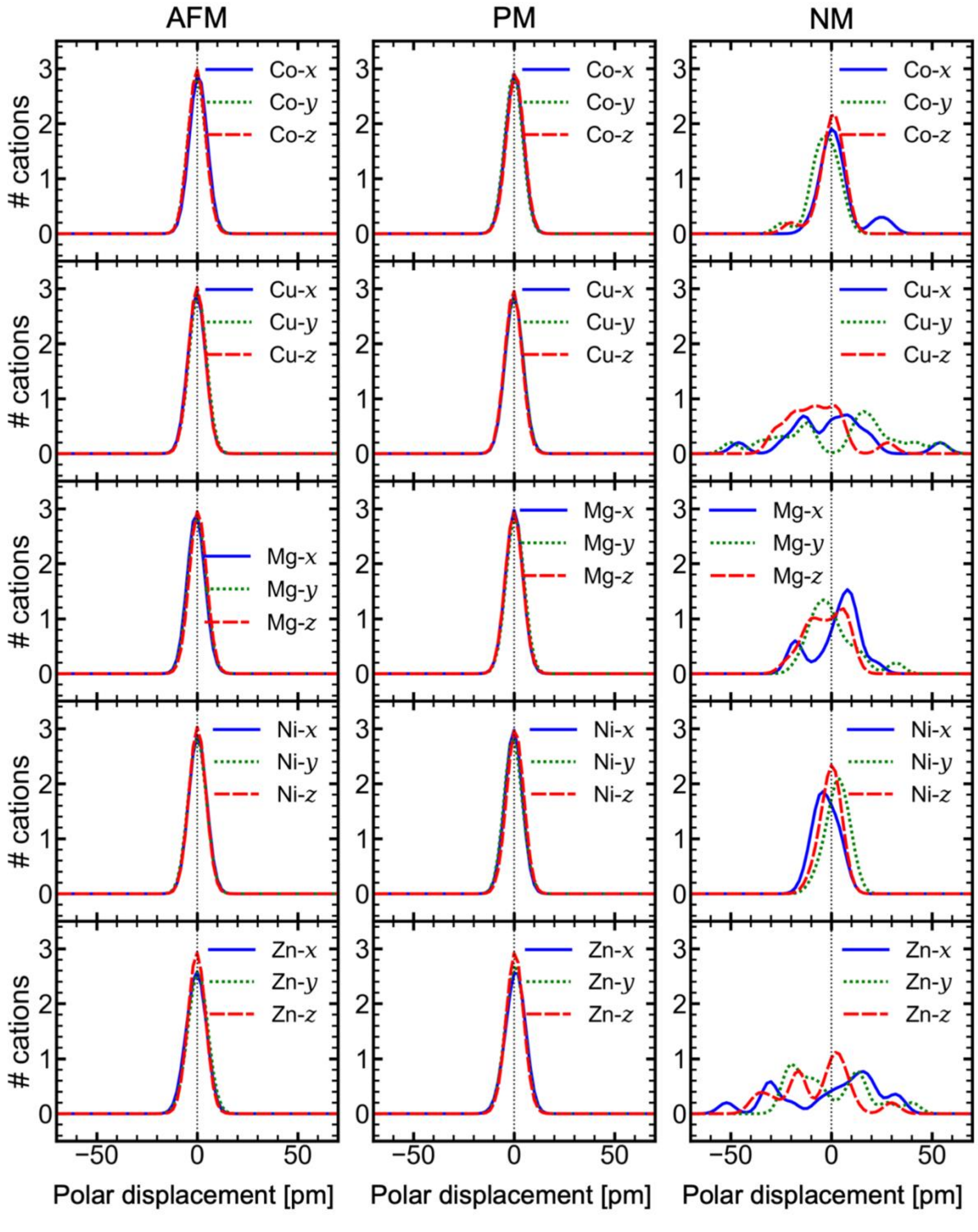

Figure S4. Distribution of polar displacements for AFM, PM and NM structures, shown in first, second and third columns, respectively. The distribution is exclusively presented for each cation separated in different rows. Cu and Zn ions in NM structure are the most distorted cations suggesting, a tendency to have local tenorite and wurtizite phases.

Table SII: Description of elastic stiffness and other derived constants. Most of the constants and their descriptions are taken from Ref. [4].

| Property | Symbol | SI Units | Formula | Description |
|---|---|---|---|---|
| Elastic stiffness tensor | $c_{ij}$ | GPa | $\sigma_i = c_{ij}\varepsilon_j$ | 6x6 symmetric matrix in Voigt notation (IEEE-format) |
| Elastic compliance tensor | $s_{ij}$ | $\mathrm{GPa}^{-1}$ | $\varepsilon_i = s_{ij}\sigma_j$ ; $s_{ij} = [c_{ij}]^{-1}$ | 6x6 symmetric matrix in Voigt notation (IEEE-format) |
| Voigt average of Bulk modulus | $K_{\mathrm{V}}$ | GPa | $9K_{\mathrm{V}} = (c_{11}+c_{22}+c_{33}) + 2(c_{12}+c_{23}+c_{31})$ | Upper bound on $K$ for a polycrystalline material |
| Voigt average of Shear modulus | $G_{\mathrm{V}}$ | GPa | $15G_{\mathrm{V}} = (c_{11}+c_{22}+c_{33}) - (c_{12}+c_{23}+c_{31}) + 3(c_{44}+c_{55}+c_{66})$ | Lower bound on $K$ for a polycrystalline material |
| Reuss average of Bulk modulus | $K_{\mathrm{R}}$ | GPa | $\frac{1}{K_{\mathrm{R}}} = (s_{11}+s_{22}+s_{33}) + 2(s_{12}+s_{23}+s_{31})$ | Upper bound on $G$ for a polycrystalline material |
| Reuss average of shear modulus | $G_{\mathrm{R}}$ | GPa | $\frac{15}{G_{\mathrm{R}}} = 4(s_{11}+s_{22}+s_{33}) - 4(s_{12}+s_{23}+s_{31}) + 3(s_{44}+s_{55}+s_{66})$ | Lower bound on $G$ for a polycrystalline material |
| Voigt-Reuss-Hill average of bulk modulus | $K_{\mathrm{VRH}}$ | GPa | $K_{\mathrm{VRH}} = \frac{K_V+K_R}{2}$ | Average of $K_{\mathrm{V}}$ , $K_{\mathrm{R}}$ |
| Voigt-Reuss-Hill average of shear modulus | $G_{\mathrm{VRH}}$ | GPa | $G_{\mathrm{VRH}} = \frac{G_{\mathrm{V}}+G_R}{2}$ | Average of $G_{\mathrm{V}}$ , $G_{\mathrm{R}}$ |
| Isotropic Young's modulus | $E$ | GPa | $\frac{1}{E} = \frac{1}{3G_{\mathrm{VRH}}} + \frac{1}{9K_{\mathrm{VRH}}}$ | Average Young's modulus for a polycrystalline material |
| Isotropic Poisson's ratio | $\nu$ | - | $\nu = \frac{1}{2}\left(1 - \frac{3G_{\mathrm{VRH}}}{3K_{\mathrm{VRH}}+G_{\mathrm{VRH}}}\right)$ | Average Poisson's for a polycrystalline material |
| Universal Anisotropic ratio | $A^{\mathrm{U}}$ | - | $A^U = 5\frac{G_{\mathrm{V}}}{G_{\mathrm{R}}} + \frac{K_{\mathrm{V}}}{K_{\mathrm{R}}} - 6$ | Universal anisotropic factor |
| Cubic anisotropic ratio | $A^{\mathrm{Z}}$ | - | $A^Z = \frac{2\,(s_{11}-s_{12})}{s_{44}} = \frac{2\,c_{44}}{c_{11}-c_{12}}$ | Anisotropic ratio for cubic single crystals |
| Young's modulus as a function of direction cosines | $E(l_1, m_1, n_1)$ | GPa | $\frac{1}{E(l_1,m_1,n_1)} = s_{11} - 2\left(s_{11} - s_{12} - \frac{1}{2}s_{44}\right)(l_1^2 l_2^2 + l_2^2 l_3^2 + l_3^2 l_1^2)$ | Young's modulus as function of direction cosines of a single crystal |

Table SIII: Values of bulk lattice constants of constituent oxides

| Compound | Phase | $a$ [Å] | $b$ [Å] | $c$ [Å] | $\alpha$ [°] | $\beta$ [°] | $\gamma$ [°] | Vol [Å$^3$] |
|---|---|---|---|---|---|---|---|---|
| MgO | Cubic-NM | 4.219 | 4.219 | 4.219 | 90.000 | 90.000 | 90.000 | 75.088 |
| CoO | Cubic-AFM | 4.135 | 4.135 | 4.135 | 90.000 | 90.000 | 90.000 | 70.692 |
| NiO | Cubic-AFM | 4.129 | 4.129 | 4.129 | 90.000 | 90.000 | 90.000 | 70.413 |
| CuO | Cubic-NM | 4.130 | 4.130 | 4.130 | 90.000 | 90.000 | 90.000 | 70.441 |
| CuO | Tenorite-NM | 4.690 | 3.330 | 5.076 | 90.000 | 100.275 | 90.000 | 78.011 |
| ZnO | Cubic-NM | 4.271 | 4.271 | 4.271 | 90.000 | 90.000 | 90.000 | 77.932 |
| ZnO | Hexagonal-NM | 3.236 | 3.236 | 5.223 | 90.000 | 90.000 | 60.000 | 47.374 |

Table SIV: Values of elastic constants, bulk modulus, shear modulus, Young's modulus, Poisson's ratio, Zener anisotropy ratio and universal anisotropy ratio for individual binary oxides and ESO. Comparison with DFT calculations and experimental measurements are also provided for reference.

| Compound | Phase | Reference | Method | $c_{11}$ / $c_{22}$ / $c_{33}$ | $c_{12}$ / $c_{23}$ / $c_{31}$ | $c_{44}$ / $c_{55}$ / $c_{66}$ | $K_{\mathrm{VRH}}$ | $G_{\mathrm{VRH}}$ | $E$ | $\nu$ | $A^{\mathrm{Z}}$ | $A^{\mathrm{U}}$ |
|---|---|---|---|---|---|---|---|---|---|---|---|---|
| | | | | [GPa] | [GPa] | [GPa] | [GPa] | [GPa] | [GPa] | - | - | - |
| MgO | Cubic-NM | [0] | DFT; U=0 | 289.53 | 86.37 | 136.10 | 154.09 | 121.05 | 287.80 | 0.19 | 1.34 | 0.10 |
| MgO | Cubic-NM | mp-1265 | DFT; U=0 | 273.00 | 91.00 | 141.00 | 151.67 | 118.30 | 281.66 | 0.19 | 1.55 | 0.23 |
| MgO | Cubic-NM | [2] | EXP at 25C | 289.00 | 88.00 | 155.00 | 155.00 | 130.29 | 305.31 | 0.17 | 1.54 | 0.23 |
| CoO | Cubic-NM | [0] | DFT; U=0 | 395.00 | 191.00 | 36.00 | 259.00 | 55.49 | 155.36 | 0.40 | 0.35 | 1.42 |
| CoO | Cubic-AFM | [0] | DFT; U=0 | 313.37 | 146.36 | 59.81 | 202.03 | 68.38 | 184.34 | 0.35 | 0.72 | 0.14 |
| CoO | Cubic-AFM | [0] | DFT; U=6 | 313.36 | 124.81 | 85.36 | 187.66 | 88.82 | 230.15 | 0.30 | 0.91 | 0.01 |
| CoO | Cubic-AFM | mp-19079 | DFT; U=3 | -42.00 | 375.00 | 60.00 | 236.00 | 38.17 | 108.65 | 0.42 | -0.29 | -6.92 |
| CoO | Cubic-AFM | [6] | DFT; U=0 | 260.30 | 163.80 | 40.10 | 195.97 | 43.18 | 120.68 | 0.40 | 0.83 | 0.04 |
| CoO | Cubic-AFM | [6] | DFT; U=8 | 212.80 | 226.70 | 130.30 | 222.07 | 28.26 | 81.32 | 0.44 | -18.75 | -24.96 |
| CoO | Cubic-AFM | [7] | EXP at 27C | 266.40 | 147.20 | 83.60 | 186.93 | 73.00 | 193.78 | 0.33 | 1.40 | 0.14 |
| NiO | Cubic-NM | [0] | DFT; U=0 | 360.00 | 205.00 | 36.00 | 256.67 | 49.21 | 138.75 | 0.41 | 0.47 | 0.74 |
| NiO | Cubic-AFM | [0] | DFT; U=0 | 389.16 | 131.56 | 43.09 | 217.43 | 68.05 | 184.86 | 0.36 | 0.34 | 1.59 |
| NiO | Cubic-AFM | [0] | DFT; U=6 | 415.72 | 117.97 | 91.69 | 217.22 | 111.45 | 285.51 | 0.28 | 0.62 | 0.29 |
| NiO | Cubic-AFM | mp-19009 | DFT; U=6 | 274.00 | 170.00 | 83.00 | 204.67 | 68.81 | 185.63 | 0.35 | 1.60 | 0.27 |
| NiO | Cubic-AFM | [8] | EXP at 0 K | 211.00 | 121.00 | 109.00 | 151.00 | 76.44 | 196.21 | 0.28 | 2.42 | 1.00 |
| NiO | Cubic-AFM | [8] | EXP at 298 K | 224.00 | 97.00 | 110.00 | 139.33 | 88.24 | 218.58 | 0.24 | 1.73 | 0.37 |
| NiO | Cubic-AFM | [9] | EXP at 298 K | 271.00 | 125.00 | 105.00 | 173.67 | 90.77 | 231.90 | 0.28 | 1.44 | 0.16 |
| NiO | Cubic-AFM | [10] | Semi empirical | 450.00 | 163.00 | 163.00 | 258.67 | 154.90 | 387.37 | 0.25 | 1.14 | 0.02 |
| NiO | Cubic-AFM | [11] | EXP at 298 K | 344.60 | 141.30 | 40.00 | 209.07 | 58.74 | 161.12 | 0.37 | 0.39 | 1.12 |
| NiO | Cubic-AFM | [6] | DFT; U=0 | 365.90 | 81.90 | 77.50 | 176.57 | 99.00 | 250.24 | 0.26 | 0.55 | 0.45 |
| NiO | Cubic-AFM | [6] | DFT; U=8 | 279.60 | 231.90 | 126.30 | 247.80 | 65.89 | 181.58 | 0.38 | 5.30 | 4.18 |
| CuO | Cubic-NM | [0] | DFT; U=0 | 254.00 | 183.00 | 50.00 | 206.67 | 43.59 | 122.18 | 0.40 | 1.41 | 0.14 |
| CuO | Cc-AFM | [0] | DFT; U=0 | 154.44 | 144.77 | 67.82 | 159.73 | 21.62 | 62.07 | 0.44 | 1.18 | 11.81 |
| | | | | 151.31 | 143.84 | 8.13 | | | | | | |
| | | | | 367.83 | 143.60 | 9.91 | | | | | | |
| CuO | Cc-AFM | [0] | DFT; U=6 | 295.68 | 152.58 | 100.65 | 140.87 | 46.77 | 126.32 | 0.35 | 1.07 | 4.12 |
| | | | | 126.74 | 85.59 | 18.35 | | | | | | |
| | | | | 327.79 | 151.57 | 56.60 | | | | | | |
| CuO | Cubic-NM | mp-14549 | DFT ; U=0 | 201.00 | 166.00 | 50.00 | 177.67 | 32.84 | 92.81 | 0.41 | 2.86 | 1.45 |
| CuO | C2/c-NM | mp-704645 | DFT ; U=0 | 202.00 | 77.00 | 13.00 | 145.81 | 25.42 | 72.06 | 0.42 | 0.16 | 5.83 |
| | | | | 202.00 | 122.00 | 13.00 | | | | | | |
| | | | | 335.00 | 122.00 | 8.00 | | | | | | |
| ZnO | Cubic-NM | [0] | DFT | 258.13 | 164.10 | 85.12 | 195.45 | 67.08 | 180.58 | 0.35 | 1.81 | 0.44 |
| ZnO | Hexagonal -NM | [0] | DFT | 204.89 | 110.98 | 39.65 | 150.07 | 40.39 | 111.19 | 0.38 | 0.85 | 0.05 |
| | | | | 204.89 | 125.58 | 36.98 | | | | | | |
| | | | | 220.11 | 125.58 | 36.98 | | | | | | |
| ZnO | Cubic-NM | mp-2229 | DFT | 233.00 | 142.00 | 83.00 | 172.33 | 65.21 | 173.72 | 0.33 | 1.82 | 0.45 |
| ZnO | Hexagonal -NM | mp-2133 | DFT | 188.00 | 109.00 | 37.00 | 129.67 | 41.27 | 111.94 | 0.36 | 0.80 | 0.13 |
| | | | | 188.00 | 92.00 | 37.00 | | | | | | |
| | | | | 205.00 | 92.00 | 39.00 | | | | | | |
| ZnO | Hexagonal -NM | [3] | DFT; U=0 | 217.00 | 121.00 | 44.00 | 152.11 | 46.53 | 126.68 | 0.36 | 0.87 | 0.03 |
| | | | | 217.00 | 117.00 | 44.00 | | | | | | |
| | | | | 225.00 | 117.00 | 44.00 | | | | | | |
| ZnO | Hexagonal -NM | [4] | Scattering; Film | 206.00 | 118.00 | 44.30 | 147.88 | 44.63 | 121.65 | 0.36 | 0.99 | 0.00 |
| | | | | 206.00 | 118.00 | 44.60 | | | | | | |
| | | | | 211.00 | 118.00 | 44.60 | | | | | | |
| ZnO | Hexagonal -NM | [5] | Bulk | 208.00 | 103.50 | 45.10 | 139.09 | 48.45 | 130.22 | 0.34 | 0.85 | 0.03 |
| | | | | 208.00 | 103.50 | 45.55 | | | | | | |
| | | | | 215.00 | 103.50 | 45.55 | | | | | | |

Antiferromagnetic structure
2.06834970389531
-0.0156487553801538 4.0787831835855171 4.0228012994594540
4.0633095795974663 -0.0165833475248264 4.0214064978945254
5.0915006300990253 5.1097927291543002 0.0420505598874117
Co Cu Mg Ni Zn O
16 16 16 16 16 80
Direct
0.2442561984898342 0.7499000992103803 0.2140036571137945
0.2360507520183028 0.0064617365703697 0.8173935606100865
0.4879149286631718 0.2519456804231878 0.8167585101756105
0.4922159139329592 0.7487582587519945 0.2142498988704796
0.4906300958967250 0.7511741997109330 0.0142469343322365
0.4939148591354566 0.9988506921855541 0.2118227741819992
0.4889907225860134 0.0029374401344873 0.6139922077153571
0.4927597850739088 0.0014585480411761 0.8116266462736156
0.7409634848341485 0.4987463371848746 0.4134181821070046
0.7381144516689575 0.5046344305194996 0.6127784691693464
0.7444097993537558 0.4979530726777940 0.0118857265100118
0.7386096210443012 0.7491074004262038 0.2122091136604899
0.7447512329058111 0.9992631345891625 0.4140429748899684
0.9961569637283066 0.2510496970952423 0.2111656779970011
0.9908669801010630 0.0029972331862956 0.6125718654155871
0.9860212658207327 0.0084775475284295 0.8125085044125969
0.2417854166978821 0.2531058542019893 0.2133499775432584
0.2438728869840783 0.2501961170974525 0.4172004149305511
0.2458760182814128 0.2440138343246153 0.0187759221970257
0.2418410744203957 0.4986143700999307 0.6170282013228992
0.2402682911812254 0.7519240301641283 0.0156915131812711
0.4919709790805127 0.4999145009770248 0.4151956110218893
0.4890624797579947 0.7529421360397089 0.6138064687683285
0.7415471976170669 0.2495808838601013 0.6163321935772647
0.7424612136161397 0.5051616427361665 0.2103798193426208
0.7436559196519186 0.0010522439629413 0.6105699457461065
0.9975093959304951 0.2476453255107209 0.0122658071700747
0.9898160375100099 0.5046040422498765 0.2152474438302305
0.9851794401894831 0.5068776686270309 0.6158705273826228
0.9913740808113753 0.5027979105872968 0.8134740688535431
0.9947865207781496 0.9963128860958580 0.4123120422346704
0.9890338651881759 0.0057062432087718 0.0117580728670113
0.2368610552602627 0.7511546670159186 0.6173014058900492
0.2475523505140577 0.9974433148969902 0.2150777788139666
0.2390770572577156 0.0048328160998740 0.6165472568493914
0.4907388579004313 0.2483258934367591 0.2145307462121493

| | | |
|---|---|---|
| 0.4925900839443965 | 0.5007402429421589 | 0.2159135416974563 |
| 0.4948074578453481 | 0.4971547374715443 | 0.6157361468892439 |
| 0.4884014478794013 | 0.7522883848462913 | 0.4154440141020402 |
| 0.7481544039173316 | 0.2500264992755930 | 0.2101483924204633 |
| 0.7350680340350907 | 0.2596895096660443 | 0.8122929087400180 |
| 0.7383497350243351 | 0.7512023050572404 | 0.4130357693240553 |
| 0.7423528132531767 | 0.0004683150913562 | 0.2108524511401706 |
| 0.7393796039159405 | 0.0019942411212373 | 0.8129297350538993 |
| 0.9908939664245158 | 0.2582249763825583 | 0.8120092665604141 |
| 0.9915174094250077 | 0.4982143097906670 | 0.0154216628575332 |
| 0.9870863936704072 | 0.7522810018036603 | 0.6125497800424683 |
| 0.9861926232603851 | 0.7556463011891420 | 0.0135277129062780 |
| 0.2408762738893682 | 0.2551833562974526 | 0.6167900429477904 |
| 0.2464844280138094 | 0.2496190489725456 | 0.8170716417804934 |
| 0.2375968136646932 | 0.5002502225483850 | 0.0177677169961205 |
| 0.2415973330346819 | 0.7482556745217107 | 0.4157334212097343 |
| 0.2441694581925211 | 0.0014447174537558 | 0.0141791541805518 |
| 0.4951565085081892 | 0.2488461499306205 | 0.6158217999770237 |
| 0.4912288640191408 | 0.2491831121645055 | 0.0147398545622153 |
| 0.4916145254993632 | 0.0003529115877741 | 0.4135450512183335 |
| 0.4919950554856935 | 0.0006679285411739 | 0.0116100784052397 |
| 0.7422895073854672 | 0.2553937228136364 | 0.0099779982473531 |
| 0.7390652541432686 | 0.5067138032297410 | 0.8102127951819947 |
| 0.7382546122655547 | 0.7539042146708391 | 0.6099323057263885 |
| 0.7389660012624042 | 0.7523493260960251 | 0.0116556068818431 |
| 0.9896524401852339 | 0.2548271769274867 | 0.6153183234536633 |
| 0.9904039919161837 | 0.7506630111740378 | 0.4140436210357504 |
| 0.9884789510238899 | 0.7532308848754304 | 0.8151866537448117 |
| 0.2401160320782626 | 0.5024957499241269 | 0.2146424665044459 |
| 0.2411533500676974 | 0.4993223240774399 | 0.4183238008763136 |
| 0.2370883695691511 | 0.5006504911432008 | 0.8172111463482320 |
| 0.2362371106266650 | 0.7555855988234323 | 0.8157677257537822 |
| 0.2418591210549770 | 0.0031233494074439 | 0.4134503101590841 |
| 0.4952997268356578 | 0.2466195164680768 | 0.4141505373610639 |
| 0.4849945332115372 | 0.5053401430982311 | 0.8174922924140779 |
| 0.4922685357509783 | 0.4995612792544241 | 0.0156096880251572 |
| 0.4874976277543401 | 0.7559480193728775 | 0.8131740907188783 |
| 0.7432824868343327 | 0.2518063719076177 | 0.4135177854461025 |
| 0.7433719069132720 | 0.7495949740812557 | 0.8119338165930383 |
| 0.7400735463659022 | 0.0074666095427978 | 0.0104492686021423 |
| 0.9952855591545697 | 0.2503744002031776 | 0.4135656410310335 |
| 0.9909068149856626 | 0.5009741284681033 | 0.4187167627650837 |
| 0.9917699172240819 | 0.7530800217877354 | 0.2130877579363548 |
| 0.9972673782522435 | 0.9984336687053237 | 0.2125198910369306 |
| 0.1293978746646999 | 0.1171865493197949 | 0.1176669910724551 |
| 0.1245635645615639 | 0.1359295487176730 | 0.3068088842170625 |
| 0.1176967854948253 | 0.1298784457084323 | 0.5171917238572226 |
| 0.1134619786482621 | 0.1358080913006145 | 0.7125941034733561 |
| 0.1144509293126882 | 0.1300513210695538 | 0.9177032829608303 |
| 0.1055605894814707 | 0.3903654973893806 | 0.1234238677245198 |
| 0.1240266300613781 | 0.3714550013764721 | 0.3291273345496383 |
| 0.1082939421591562 | 0.3821111529126911 | 0.5306567751026209 |
| 0.1270200824283553 | 0.3770439690309017 | 0.7208494200975154 |
| 0.1284253723968125 | 0.3595518004952914 | 0.9266528958751553 |
| 0.1157198282269246 | 0.6286498254453448 | 0.1117636074886632 |
| 0.1143149044371814 | 0.6310922302611791 | 0.3161182735115586 |
| 0.1142797427245552 | 0.6249932048487458 | 0.5182019322794978 |
| 0.1043327359036901 | 0.6299713508580765 | 0.7224474900770015 |
| 0.1119329763878883 | 0.6287882315106771 | 0.9167265499639590 |
| 0.1294236627529975 | 0.8685873546053786 | 0.1131522765378549 |
| 0.1139288991322607 | 0.8775835860101616 | 0.3225714428217864 |
| 0.1204284145655273 | 0.8719328085001397 | 0.5153585197889655 |
| 0.1114027237431073 | 0.8798601633076687 | 0.7141399960445695 |
| 0.0992683631578127 | 0.8927528561735459 | 0.9150611053786786 |
| 0.3720708259646938 | 0.1235336786426656 | 0.1064551316566510 |
| 0.3765253104902320 | 0.1157908612675126 | 0.3111877664101699 |
| 0.3614706403012962 | 0.1298874673155480 | 0.5127645018281702 |
| 0.3649951235732046 | 0.1324487813436008 | 0.7115887297897862 |
| 0.3731272754561271 | 0.1167612423802889 | 0.9101004451632703 |
| 0.3619918435914484 | 0.3685046948426968 | 0.1169823797807969 |
| 0.3687931640870947 | 0.3704854954472592 | 0.3166902814495732 |
| 0.3820797477198656 | 0.3709560569926799 | 0.5105187626352968 |
| 0.3677595680622668 | 0.3755400107537693 | 0.7146544793487166 |
| 0.3603824095363384 | 0.3737482683578580 | 0.9229169823608786 |
| 0.3760939258696372 | 0.6151920487321189 | 0.1174356853917541 |
| 0.3605971844466123 | 0.6307248391583286 | 0.3124416171554217 |
| 0.3607297512020104 | 0.6195063757395638 | 0.5194857734423476 |
| 0.3654092531537809 | 0.6321626777433462 | 0.7106558293635463 |
| 0.3546660094705892 | 0.6300379009575156 | 0.9209169300747637 |
| 0.3689412483168484 | 0.8749207305812712 | 0.1133565482580647 |
| 0.3657331270087445 | 0.8720936766474074 | 0.3159513087257767 |
| 0.3691089367912790 | 0.8725920148703091 | 0.5184278088431987 |
| 0.3554391937169512 | 0.8869657699803700 | 0.7197559063609434 |
| 0.3612783692151943 | 0.8751285725604168 | 0.9179372119288897 |
| 0.6173370501818948 | 0.1303643307280688 | 0.1114383699858614 |
| 0.6201906242868532 | 0.1234140416580705 | 0.3127852665616948 |
| 0.6195561104442923 | 0.1227170668925856 | 0.5165107547012177 |
| 0.6160740816078490 | 0.1335698582714853 | 0.7139876441736830 |
| 0.6131047009925388 | 0.1287169762873678 | 0.9131816443710885 |
| 0.6210052678277307 | 0.3701888758629977 | 0.1157519145841461 |
| 0.6233369311481816 | 0.3825702816855308 | 0.3082840675069251 |
| 0.6261953014884311 | 0.3630167870078789 | 0.5227554553604551 |
| 0.6070928215559338 | 0.3849312476996091 | 0.7209100754878320 |
| 0.6145395835125894 | 0.3787646467665668 | 0.9165918725366388 |
| 0.6217692730926999 | 0.6220963695068716 | 0.1133922006210151 |
| 0.6033605672766028 | 0.6251709783265347 | 0.3234356884021906 |
| 0.6193677851220297 | 0.6206368442681763 | 0.5090772908233721 |
| 0.6098516613045533 | 0.6363010421275138 | 0.7104845754787223 |
| 0.6166303413998038 | 0.6289643238435392 | 0.9099988030964924 |
| 0.6155244297526860 | 0.8755204806660654 | 0.1122235187696031 |
| 0.6179257899787429 | 0.8745795217921473 | 0.3136172000114312 |
| 0.6082929608364718 | 0.8764302338945630 | 0.5151533027338239 |
| 0.6201651784109535 | 0.8844525586023096 | 0.7070390040823759 |
| 0.6161453445309899 | 0.8754820399999614 | 0.9153637863890425 |
| 0.8715983714976826 | 0.1264641437829023 | 0.1043888600566885 |
| 0.8687270165757058 | 0.1277217240573824 | 0.3092252692074821 |
| 0.8781709644876655 | 0.1080065258497098 | 0.5060131035133241 |
| 0.8617771785298068 | 0.1329432577508139 | 0.7100849745017184 |
| 0.8586156427167601 | 0.1408318185259576 | 0.9066999885083727 |
| 0.8793104231577999 | 0.3645965687722514 | 0.1047419918290257 |
| 0.8666843254008352 | 0.3827524572407428 | 0.3081556873517650 |
| 0.8646677335803946 | 0.3681906128811431 | 0.5186407986973929 |
| 0.8641232089169809 | 0.3835131402133085 | 0.7111605321089327 |
| 0.8636030407062575 | 0.3896668370308159 | 0.9020167438283916 |
| 0.8549469590745554 | 0.6309770023062985 | 0.1089747737708827 |
| 0.8679239051258599 | 0.6237236345755048 | 0.3070972180218200 |
| 0.8588025085081411 | 0.6341038794030898 | 0.5064601158510520 |
| 0.8592672651297362 | 0.6308254372595747 | 0.7041498362667062 |
| 0.8714588168579945 | 0.6227222173132780 | 0.9112485573464038 |
| 0.8656847066061409 | 0.8788542023009839 | 0.1116291899098924 |
| 0.8666085807298465 | 0.8773766595870746 | 0.3115397358488662 |
| 0.8670773860160256 | 0.8662810084878368 | 0.5103681560234817 |
| 0.8647876781078201 | 0.8784052727759220 | 0.7114484565538837 |
| 0.8639344826638936 | 0.8758470304244632 | 0.9110530156593771 |

```
Nonmagnetic structure
   2.06834970389531
    -0.0455177428863204    4.0101671985157870    3.8511751058593990
     4.3729330484359918   -0.0377345323420459    3.8384821987986641
     5.4395208200384060    4.9714914442384082    0.0517107767211846
   Co   Cu   Mg   Ni   Zn   O
    16    16    16    16    16    80
Direct
  0.2514478770264674  0.7632780023107574  0.2157783715342346
  0.2642400710547655  0.9792424122228263  0.8160127890221496
  0.4974422095842841  0.2320339749745155  0.8187831930067876
  0.4988052571685350  0.7547075070621371  0.2149544027806961
  0.4888125411060676  0.7579770108257146  0.0226283875526349
  0.5133754076474499  0.9861990613521504  0.2125410017226721
  0.5085406037519621  0.9847647645402874  0.6189868046892569
  0.5056774026516354  0.9867688884670605  0.8131610135914791
  0.7118054684552496  0.5089124809174042  0.4225273989584281
  0.7362447871912272  0.4873853128595508  0.6116897260433105
  0.7413827203818676  0.5024190926786459  0.0165561587129147
  0.7384968978705644  0.7548875406217076  0.2185918908384202
  0.7447014114239650  0.9930084131778234  0.4118591255129604
  0.0025645225773429  0.2598522535941452  0.1982007314172147
  0.0006509835173583  0.0081990175221799  0.6155658367932895
  0.0106176980718661  0.9872218571765844  0.8084484678743374
  0.2376604300603722  0.2677338186513615  0.2087139268364883
  0.2297472604895231  0.2614366977011989  0.4170633390284171
  0.2388865157407057  0.2693276825683190  0.0001421771840597
  0.1692099015270775  0.5118945699526138  0.6146680433579106
  0.2409376691344157  0.7631183462232707  0.0167275028838460
  0.3995366550364003  0.5154877609068281  0.4496848655787243
  0.4931196260422249  0.7351666983832458  0.6201641140379061
  0.7392746264963053  0.2350245273789648  0.5953000641589313
  0.7690678379015585  0.4894651319282202  0.2076302787274963
  0.7414953473601030  0.0222938875846800  0.5950450457427238
  0.9914152837975088  0.2516422458497585  0.0055518436423921
  0.9789962551923846  0.5162581719777294  0.2077334344008396
  0.9499910527972685  0.5126144931469386  0.6321194557974468
  0.9858056042975873  0.4922496887233948  0.8000364699732015
  0.9240896183082795  0.0798759393039106  0.4155234957155655
  0.9857829014210203  0.0445052448612466  0.0095848707363700
  0.2348191886191480  0.7586232308764524  0.6114143310432258
  0.2528151717328135  0.0296679221984207  0.2074989502015953
  0.2579989379478896  0.9904358249707923  0.6178802577457455
  0.4835508196635162  0.2494314206779496  0.2170257105908671
```

| | | |
|---|---|---|
| 0.4908692284632868 | 0.5045473046947089 | 0.2217178711402146 |
| 0.4722160868774825 | 0.4940823429650882 | 0.6202049993459803 |
| 0.4865752139466494 | 0.7536081139015336 | 0.4163806673114564 |
| 0.7406188918861409 | 0.2537974870807066 | 0.2140029270689841 |
| 0.7425529946588517 | 0.2443032323233232 | 0.8096576449035896 |
| 0.7422248368796053 | 0.7410295817804302 | 0.4152406123510922 |
| 0.7530690310301581 | 0.9999687286602109 | 0.2106801043183009 |
| 0.7552013189849088 | 0.9972229260651128 | 0.7990230684473048 |
| 0.0151696851227218 | 0.2325945367239813 | 0.7921479438096921 |
| 0.9944224567877494 | 0.5031337094781768 | 0.9999801565133228 |
| 0.9782394800765625 | 0.7877441628037729 | 0.6017822119703584 |
| 0.0011490215961556 | 0.7503041482241865 | 0.0167830181719308 |
| 0.2592586881794981 | 0.2270980856418022 | 0.6192445050445580 |
| 0.2625937547435115 | 0.2269339987264498 | 0.8105359342372868 |
| 0.2395713336825686 | 0.4906398239717549 | 0.0227354164588319 |
| 0.2438097835205322 | 0.7722771757579748 | 0.4066862169990413 |
| 0.2580496354774732 | 0.9998966803006606 | 0.0080206086556508 |
| 0.4902163395553979 | 0.2402987016885989 | 0.6226976491781872 |
| 0.5118693811143634 | 0.2294280277142977 | 0.0129489393360370 |
| 0.4997860338822517 | 0.9896213227234254 | 0.4157263621997311 |
| 0.5123950051694708 | 0.9920202315548137 | 0.0074106250302738 |
| 0.7395247522406495 | 0.2587156290886243 | 0.0103159469679665 |
| 0.7286220044175321 | 0.5037149477953257 | 0.8129377957304759 |
| 0.7350599946905756 | 0.7420957706756575 | 0.6103206048207859 |
| 0.7399777599277870 | 0.7619722995494779 | 0.0143069708563521 |
| 0.9991162803118911 | 0.2460890610779816 | 0.6086596173220526 |
| 0.9814091277340384 | 0.7691277776886681 | 0.4152187615642908 |
| 0.9867679553685670 | 0.7518377316217928 | 0.8155355654774267 |
| 0.2436906387323993 | 0.5044079743912236 | 0.2049008544362249 |
| 0.1957102557200616 | 0.5226928267339745 | 0.4372031542959795 |
| 0.2182739275864855 | 0.4967317365076120 | 0.7970120770448345 |
| 0.2482533754343524 | 0.7316112544880288 | 0.8190213478821243 |
| 0.2271895824014980 | 0.0376414910321882 | 0.4140465463772294 |
| 0.4771256776826079 | 0.2504223198207132 | 0.4137803717977018 |
| 0.4776135983864337 | 0.4869022216929545 | 0.8146791493011016 |
| 0.5088578622102122 | 0.4847785998185427 | 0.0095185264266688 |
| 0.4947984431425712 | 0.7342552722706581 | 0.8269586369757700 |
| 0.6867991802231983 | 0.2797929160085541 | 0.4299536905376316 |
| 0.7541040767129336 | 0.7448518804201583 | 0.8072657154755718 |
| 0.7550031977871673 | 0.0005015853772833 | 0.0119542319711365 |
| 0.9888019877165611 | 0.2804843212928638 | 0.3893103164851748 |
| 0.9752574714586112 | 0.4966749714761410 | 0.4234402799124605 |
| 0.0025058987685428 | 0.7572883894357181 | 0.2100560364238532 |
| 0.0138942741894006 | 0.9983923751653160 | 0.2201802928515877 |
| 0.1148973436654900 | 0.1539154383033571 | 0.1220190760129944 |
| 0.1187651834470777 | 0.1501722561848193 | 0.2931219372431292 |
| 0.1220744920212751 | 0.1260009622922589 | 0.5366942862251439 |
| 0.1506688983822655 | 0.0926989276620271 | 0.7262528379780087 |
| 0.1698382397479659 | 0.0855321254136762 | 0.8983008682729745 |
| 0.0832593986064440 | 0.4061895066230696 | 0.1207605178728500 |
| 0.0917043085100814 | 0.4106294723896370 | 0.3219820448545141 |
| 0.1053490219865675 | 0.3767189220573097 | 0.5178094945576832 |
| 0.1638163519872003 | 0.3263087601890767 | 0.7113666779113664 |
| 0.0894118736598708 | 0.3898691750931997 | 0.9150380191520304 |
| 0.1074613136178214 | 0.6501363708589390 | 0.1241255971194723 |
| 0.1063687968435371 | 0.6722707347377130 | 0.3210921920557787 |
| 0.1010646722801442 | 0.6843043380001620 | 0.4995495125057414 |
| 0.0785870988891567 | 0.6626564322110143 | 0.7217451121164192 |
| 0.1241261592744451 | 0.6141161772865987 | 0.9388743323414087 |
| 0.1325718731086124 | 0.8953608349972062 | 0.1259072712535929 |
| 0.1698037291867980 | 0.8727742933040290 | 0.2924138422030593 |
| 0.1057704682510071 | 0.9139064626140795 | 0.5248964808203688 |
| 0.1055630817972068 | 0.8775369739273337 | 0.7223144508563298 |
| 0.1332877410039747 | 0.8728123382814980 | 0.8958493786777821 |
| 0.4271164313712970 | 0.0874582598625412 | 0.1009210639292903 |
| 0.4088378137449433 | 0.0785334025085320 | 0.3093259863192823 |
| 0.3796981594641274 | 0.1108976666952588 | 0.5433665726760195 |
| 0.4174769087815672 | 0.0786923252137528 | 0.7094623089282892 |
| 0.3851954057899647 | 0.1024042336236286 | 0.8888907042692925 |
| 0.3559954683813854 | 0.3865810667708722 | 0.1155563381910791 |
| 0.3693420789920439 | 0.3665440095073996 | 0.3046709776213488 |
| 0.3864380931532802 | 0.3279076817561936 | 0.5310593481986823 |
| 0.3638657426966042 | 0.3534828192630047 | 0.7441941013678885 |
| 0.4108112509652226 | 0.3122629546954654 | 0.9196063162035936 |
| 0.3545911524885321 | 0.6527395707943840 | 0.1313245762210956 |
| 0.3798978603271730 | 0.6370530663138488 | 0.2964811295876697 |
| 0.3364904899229374 | 0.6536757565257777 | 0.5174949195490276 |
| 0.3475232376959863 | 0.6171796100080862 | 0.7202619057710822 |
| 0.3880037483922619 | 0.5872815310037850 | 0.9107733799980324 |
| 0.3858910114916744 | 0.8769290573281452 | 0.0985569356122027 |
| 0.3928892797779235 | 0.8646956481137048 | 0.2967559429985693 |
| 0.3722463108929862 | 0.8826883686253918 | 0.5156209374799656 |
| 0.3706794744026279 | 0.8751919893853679 | 0.7269363477427087 |
| 0.3927429872156125 | 0.8578636356571198 | 0.9085930566853995 |
| 0.6216897815217007 | 0.1297225389956660 | 0.1089494393459291 |
| 0.6276339685247708 | 0.1005723312583141 | 0.3163155771421640 |
| 0.6386428351391540 | 0.0926228309949333 | 0.4994838504584451 |
| 0.5940387203222084 | 0.1406729976237308 | 0.7231958445740559 |
| 0.6359885124140443 | 0.1015689254540469 | 0.8963937377701199 |
| 0.6067510814867568 | 0.3968441681223448 | 0.1208752145792690 |
| 0.6174969669976678 | 0.3575787594847676 | 0.3092229654952837 |
| 0.5727010525825345 | 0.3987472603172736 | 0.5189794184977077 |
| 0.6005038686636675 | 0.3733970312381414 | 0.7258314763203431 |
| 0.6378597240563033 | 0.3535893706401189 | 0.9129551993736958 |
| 0.5986585292032409 | 0.6678569372172142 | 0.1163104581898534 |
| 0.6063789935901991 | 0.6501510447212173 | 0.3124526396952674 |
| 0.6131570866516219 | 0.6113980600473629 | 0.5164626519425756 |
| 0.6429888200078286 | 0.5953490434676432 | 0.7037875843554362 |
| 0.6011547040612221 | 0.6382669531200067 | 0.9501063775798626 |
| 0.6214245321421606 | 0.8784353162883585 | 0.1346954632990925 |
| 0.6335926127660783 | 0.8630479733526910 | 0.3034325028594222 |
| 0.6113218614444008 | 0.8841526806433787 | 0.5192190675747195 |
| 0.6096399013584521 | 0.8735284542480016 | 0.7322562946757333 |
| 0.6062601280954000 | 0.8830536394505515 | 0.9240639954003975 |
| 0.8474863082282706 | 0.1555638110893591 | 0.1025804724399952 |
| 0.8766111905721057 | 0.1155504551844406 | 0.2985511348617162 |
| 0.9068168747170007 | 0.1350767121500371 | 0.5167724619384877 |
| 0.9043932426924952 | 0.1054488353760931 | 0.7090305398411921 |
| 0.8958498718806220 | 0.0990516937453206 | 0.8879299387248739 |
| 0.8603765413245730 | 0.3870744670523644 | 0.1127527587349445 |
| 0.8163472586790291 | 0.3960923207746058 | 0.3483875651290785 |
| 0.8113508242072052 | 0.4002983990005529 | 0.5208745886709454 |
| 0.8777690876964610 | 0.3505862716326456 | 0.6895036383232381 |
| 0.8116250866955683 | 0.4158419040946464 | 0.9224988982079156 |
| 0.8328480466689886 | 0.6698838030992236 | 0.1169363923048013 |
| 0.8369748290110677 | 0.6307761039202537 | 0.3213395040446966 |
| 0.8547713812790184 | 0.6027969817565774 | 0.5132478418567247 |
| 0.8325169136823095 | 0.6375133371854061 | 0.7137078168319189 |
| 0.8566683048436718 | 0.6356972594864082 | 0.9149398246331536 |
| 0.8650216482760225 | 0.8769827987850205 | 0.1229359940544831 |
| 0.8590636408781253 | 0.8794770240445456 | 0.3274011007461251 |
| 0.8555436950770758 | 0.8579008092986913 | 0.5071442161995473 |
| 0.8572768675934481 | 0.9067767189076436 | 0.6929027204300362 |
| 0.8878117742298244 | 0.8598169619916469 | 0.8995989460039761 |

```
Paramagnetic Structure 1
  2.06834970389531
   -0.0165183403311863    4.0630222025014726    4.0244383211151842
    4.0759443716687453   -0.0158266691114973    4.0215115404239397
    5.1088546120140004    5.0896957017060780    0.0413904824766581
  Co   Cu   Mg   Ni   Zn   O
   16    16    16    16    16    80
Direct
  0.2476298353337764  0.7463520950063931  0.2165146145896088
  0.2349566120771360  0.0072990922271562  0.8179152421699235
  0.4873887310833285  0.2536015118069990  0.8159336730236665
  0.4933397844366604  0.7482040801131252  0.2148519306719031
  0.4913423102248607  0.7502107470085441  0.0149815056599562
  0.4941234218889643  0.9986170800745463  0.2121714828116264
  0.4883025861547536  0.0035834357666300  0.6137950124056788
  0.4915937050038822  0.0023052720432424  0.8112663667691168
  0.7405065372998191  0.4994456002492447  0.4131592539323020
  0.7375061662704893  0.5054177106278126  0.6124988914715932
  0.7457403959420985  0.4968459760464974  0.0118395587464143
  0.7387859416765166  0.7481742312238326  0.2131434378615351
  0.7452303777594966  0.9961830168636402  0.4140421170420944
  0.9980590179421774  0.2507349377272635  0.2108011760961911
  0.9884658361615730  0.0063120897564049  0.6120985771442867
  0.9848300861360528  0.0101260070545153  0.8119952668659761
  0.2436327548640351  0.2537240001332068  0.2122940822207834
  0.2440465541294109  0.2500575824514703  0.4175473362029007
  0.2468599911549947  0.2437653255748788  0.0175451198363204
  0.2399740938413099  0.5014099139162181  0.6174481958885323
  0.2402608603688457  0.7518617917082409  0.0160986709302199
  0.4913365746261550  0.4991669621034875  0.4169055098682340
  0.4879850682729874  0.7542980063670608  0.6137326806349005
  0.7378270819144650  0.2526281019954416  0.6153803040703519
  0.7442858900948235  0.5023247610876296  0.2106505540182405
  0.7431086913310182  0.0015027570591604  0.6100540694358522
  0.9993089412505961  0.2468395685399346  0.0113056612841456
  0.9907643810402536  0.5038709338639472  0.2157189166827046
  0.9848776282009043  0.5067655706775402  0.6166436706299465
  0.9906175794936022  0.5034054973007297  0.8132842141122293
  0.9947881271195178  0.9956003645281687  0.4123270991832582
  0.9892716712204939  0.0047172306248385  0.0119618522171760
  0.2339548825392832  0.7543335045214600  0.6181430460573839
  0.2481501881327795  0.9975971041540994  0.2152134140767177
  0.2380418823447124  0.0059046119279059  0.6166609678845794
  0.4922600577813364  0.2477458754176716  0.2137141389931312
```

| | | |
|---|---|---|
| 0.4975355533115168 | 0.4963188071117899 | 0.2162290764974107 |
| 0.4910016454910296 | 0.5011889757362357 | 0.6155070475975220 |
| 0.4873750589869212 | 0.7513029749689221 | 0.4163528801083143 |
| 0.7518326809200960 | 0.2465751960794165 | 0.2089040171140153 |
| 0.7339906439076621 | 0.2604477520200905 | 0.8112796983820332 |
| 0.7376949653827215 | 0.7510841412844887 | 0.4131530007426150 |
| 0.7430460599255275 | 0.9991294087168965 | 0.2106411536087064 |
| 0.7390703773711097 | 0.0024280501762339 | 0.8119859636155634 |
| 0.9904359366205046 | 0.2585955636992013 | 0.8116057530568025 |
| 0.9918824456019882 | 0.4984312196016181 | 0.0151489190573449 |
| 0.9850338919682445 | 0.7534221200654222 | 0.6128678142954460 |
| 0.9858743348430964 | 0.7556707254393908 | 0.0138120751724099 |
| 0.2412632148292202 | 0.2558424417629362 | 0.6162882764346779 |
| 0.2452256494590924 | 0.2504947395465385 | 0.8171837269691966 |
| 0.2380782513174586 | 0.5009831103215343 | 0.0173795099599181 |
| 0.2421393879048460 | 0.7477240544387678 | 0.4178915196738588 |
| 0.2450421681578892 | -0.0004610094601678 | 0.0146584590644234 |
| 0.4923295379115776 | 0.2504964302903568 | 0.6156443336042080 |
| 0.4916913624794987 | 0.2494685737208179 | 0.0143381477569777 |
| 0.4917873394423386 | 0.0006105966936522 | 0.4129719080299463 |
| 0.4924062123496218 | 0.0004821295817052 | 0.0110795396536250 |
| 0.7425791987709347 | 0.2548927107222206 | 0.0090982279034743 |
| 0.7384048523001796 | 0.5065614706446722 | 0.8105215270551475 |
| 0.7376462439383764 | 0.7532172318499349 | 0.6106479400632387 |
| 0.7383751612872843 | 0.7523611676313746 | 0.0126305338334680 |
| 0.9888364373904853 | 0.2554925186070671 | 0.6152999025235880 |
| 0.9909967159380888 | 0.7493404847646193 | 0.4147223933417750 |
| 0.9869110658362795 | 0.7542976410795453 | 0.8159451613511564 |
| 0.2411348989720721 | 0.5012806154584944 | 0.2154561523529967 |
| 0.2409841825530444 | 0.4976697198152730 | 0.4198665745905508 |
| 0.2356582364926302 | 0.5029891477995515 | 0.8166108492058763 |
| 0.2353864893721639 | 0.7568630387154217 | 0.8164698163457991 |
| 0.2418105814459335 | 0.0042900610430234 | 0.4129655871721098 |
| 0.4976466952364370 | 0.2469456463391396 | 0.4124879164418658 |
| 0.4850205034148729 | 0.5059312106791598 | 0.8168401902275493 |
| 0.4922333164401814 | 0.5000495385965383 | 0.0158831382940702 |
| 0.4868723241468277 | 0.7560030357235636 | 0.8134119671150734 |
| 0.7435791084442920 | 0.2507164989671311 | 0.4120398967957576 |
| 0.7428629237811685 | 0.7496658156540114 | 0.8121615396040668 |
| 0.7407785228909471 | 0.0071209165906228 | 0.0098644530488854 |
| 0.9955022036837955 | 0.2510570749850048 | 0.4118464918451317 |
| 0.9907502696068848 | 0.5008402288414637 | 0.4190990234457247 |
| 0.9938041103083041 | 0.7519828560591207 | 0.2133601517062031 |
| 0.9991514734287575 | 0.9965454059264623 | 0.2115683396196896 |
| 0.1317099731496806 | 0.1168642818373943 | 0.1161443902647177 |
| 0.1278846701380421 | 0.1370198069891841 | 0.3049494033880692 |
| 0.1171370940445723 | 0.1312715292788217 | 0.5167132923494377 |
| 0.1130340112768045 | 0.1364397816569464 | 0.7127126585987092 |
| 0.1137811162048136 | 0.1300243135235276 | 0.9173223122794115 |
| 0.1064142007028982 | 0.3905937098637340 | 0.1234989916133599 |
| 0.1221521272711488 | 0.3733954665323002 | 0.3297134758933271 |
| 0.1068830052230714 | 0.3825649223758368 | 0.5312194082992492 |
| 0.1262971069318707 | 0.3772845671725732 | 0.7208246528601152 |
| 0.1289218323711947 | 0.3609256661849933 | 0.9252097622118107 |
| 0.1173058989639288 | 0.6285083961945045 | 0.1118105860437368 |
| 0.1166811667558442 | 0.6295810129696608 | 0.3175569446226453 |
| 0.1148323748404337 | 0.6242154730004981 | 0.5199799463683992 |
| 0.1008117639328083 | 0.6326438015455159 | 0.7238313149250176 |
| 0.1116815114907041 | 0.6291232982734236 | 0.9174464175915510 |
| 0.1300953984507787 | 0.8675387993380219 | 0.1148810929095298 |
| 0.1179202106401441 | 0.8735892945950668 | 0.3216592097296455 |
| 0.1148724182467858 | 0.8773243275518759 | 0.5132886299846779 |
| 0.1094331715326730 | 0.8820829892007849 | 0.7137798160025166 |
| 0.0995420040078840 | 0.8927121665965265 | 0.9146411648895342 |
| 0.3739002016398064 | 0.1225478855364069 | 0.1054050082538576 |
| 0.3771543675968094 | 0.1169060205217386 | 0.3095085970364116 |
| 0.3623502508770747 | 0.1295993397527137 | 0.5128091725498612 |
| 0.3645533686723279 | 0.1325214082621107 | 0.7122075039198514 |
| 0.3719672384765045 | 0.1142147372364636 | 0.9101220270462463 |
| 0.3631411230645445 | 0.3692880564779190 | 0.1159626247465933 |
| 0.3701827842733206 | 0.3692828292750639 | 0.3159917361471746 |
| 0.3803116948579746 | 0.3692462088606365 | 0.5107375122421577 |
| 0.3658083561603269 | 0.3773250774686256 | 0.7144598516067067 |
| 0.3606140257659523 | 0.3730715732890058 | 0.9221850567225154 |
| 0.3756624613979167 | 0.6159413148834993 | 0.1211243444471663 |
| 0.3693823352822215 | 0.6211460450082337 | 0.3195569453193671 |
| 0.3526235789664159 | 0.6280050667407038 | 0.5251122887158827 |
| 0.3633951255686987 | 0.6341547338455509 | 0.7125295834761561 |
| 0.3558388611756256 | 0.6289183671444935 | 0.9216281059606226 |
| 0.3707105035582640 | 0.8730738464756793 | 0.1146786115400319 |
| 0.3659384871682510 | 0.8716302161836671 | 0.3164580341212427 |
| 0.3672896126279561 | 0.8755700300747228 | 0.5179494065184639 |
| 0.3546692261776142 | 0.8880017225607698 | 0.7202055874919294 |
| 0.3604294492187978 | 0.8738127397879599 | 0.9192201247327801 |
| 0.6179743788079810 | 0.1298200495655874 | 0.1108216037832920 |
| 0.6222677082757092 | 0.1210821250940522 | 0.3122573910442394 |
| 0.6204057335794643 | 0.1214650235958264 | 0.5174035152205806 |
| 0.6134159416599361 | 0.1394945426823264 | 0.7136053565858880 |
| 0.6117591958213113 | 0.1295079399956573 | 0.9120546565754617 |
| 0.6218968406596384 | 0.3694703567764984 | 0.1139341623685766 |
| 0.6336767074743209 | 0.3727633469414717 | 0.3025867385711953 |
| 0.6160070428104614 | 0.3721432588093784 | 0.5170838845270253 |
| 0.6078530714344169 | 0.3884530832140705 | 0.7179613312525647 |
| 0.6144902840207018 | 0.3786301575836890 | 0.9157178979366678 |
| 0.6229998802673107 | 0.6210315315142468 | 0.1136305844718328 |
| 0.6050412486662738 | 0.6252427898230803 | 0.3229084298159350 |
| 0.6182825397692501 | 0.6217311994687670 | 0.5091168487363851 |
| 0.6096963717672897 | 0.6372535048177830 | 0.7106429138956943 |
| 0.6162609329646062 | 0.6283200800290693 | 0.9108672960129317 |
| 0.6164506111129735 | 0.8750379335822961 | 0.1122812419367404 |
| 0.6176467324468055 | 0.8742194800306234 | 0.3135677233355409 |
| 0.6081952990178956 | 0.8762873473287673 | 0.5152016784205377 |
| 0.6181082912609814 | 0.8851430086256641 | 0.7086531571166422 |
| 0.6155526011365934 | 0.8761148064236111 | 0.9150654483208597 |
| 0.8723455210946972 | 0.1246221923139178 | 0.1046952077583147 |
| 0.8756688407218051 | 0.1207896963037545 | 0.3105715123638838 |
| 0.8765392929026480 | 0.1108324500154454 | 0.5049159216903939 |
| 0.8613330530936176 | 0.1332674589151458 | 0.7098294736060300 |
| 0.8579705014688709 | 0.1412513221865783 | 0.9057645229087216 |
| 0.8825063039417642 | 0.3620865010404055 | 0.1037609548938487 |
| 0.8678876270587380 | 0.3815408509206759 | 0.3085259036622824 |
| 0.8620242527673665 | 0.3674247945435256 | 0.5184939449729707 |
| 0.8637837571639456 | 0.3836958127319833 | 0.7110175919448855 |
| 0.8641559764278577 | 0.3894139170861719 | 0.9018070368488608 |
| 0.8555073133663217 | 0.6300499710193036 | 0.1098195472088848 |
| 0.8685929026463274 | 0.6233295032156614 | 0.3075291824083698 |
| 0.8587464348426237 | 0.6326914843249263 | 0.5069611698197813 |
| 0.8587784114629602 | 0.6303556326736474 | 0.7050654247744286 |
| 0.8706198163188619 | 0.6232230544752476 | 0.9117263920849427 |
| 0.8667340002851845 | 0.8778726009629053 | 0.1113586623445050 |
| 0.8668943374289209 | 0.8754074192949373 | 0.3118826538436882 |
| 0.8654116066704244 | 0.8631572116564815 | 0.5114203143455438 |
| 0.8616679756987152 | 0.8814935592743457 | 0.7100422975700371 |
| 0.8630721830104181 | 0.8764205029567638 | 0.9110269691820568 |

Paramagnetic Structure 2

```
   2.06834970389531
    -0.0147477094679292    4.1170937613323639    4.0332907257713719
     4.0176205516601016   -0.0162330826188179    4.0320383335978001
     5.0376347996959865    5.1613151358529992    0.0395839507541929
   Co   Cu   Mg   Ni   Zn   O
    16    16    16    16    16    80
Direct
  0.2463067653278696  0.0072973863077450  0.0026773960727196
  0.9992406976523245  0.5031420423741757  0.9980128791272406
  0.4934469357095189  0.2551005990304080  0.9980449716458250
  0.9976850803460728  0.7515740032813821  0.2015897394519604
  0.7489218186549745  0.7521246081153659  0.2020019958740401
  0.2531432046563163  0.5017553086543440  0.1949957687897238
  0.0030757565682600  0.2536310187649662  0.1946496530788446
  0.5009968749717399  0.5004324936784964  0.3990375979276621
  0.7512482601647837  0.2480975699863157  0.3975865062000820
  0.2537773274041492  0.2483521970881200  0.3964668356625747
  0.0032463511380192  0.2492957778124109  0.5948212162302632
  0.9962406494960604  0.0025574586279877  0.8001397021399645
  0.4944327565038119 -0.0015137862880741  0.8067742965553882
  0.2508411001675297  0.7511053926776393  0.8031434169445637
  0.9983581474274215  0.5028890769393551  0.7976642210961664
  0.2482647212276700  0.5042978204692978  0.7984620093795493
  0.4926916877009998  0.0061723646094217  0.0018431212012201
  0.2490563304083084  0.7536780286374845  0.0034804908864288
  0.7457961536107047  0.2518356355193667  0.9989454575742126
  0.4995627110888207  0.0003112644368789  0.2007545731363824
  0.2505872988682198  0.0033400880438539  0.1969291229260578
  0.4996672985673446  0.2497610125859238  0.1960170890384128
  0.7531880324199933  0.9923076855559722  0.4044621742177206
  0.5031079443103442  0.9940061429570178  0.4050374059064120
  0.0047305838551003  0.5003259397839823  0.3972489335049771
  0.2512616366921507  0.9991986709371423  0.5996164480971909
  0.5020554075366706  0.7425892484951471  0.6062744597878756
  0.7543483047148015  0.2443458501270387  0.5980578587289190
  0.7434197237074075  0.0020788022564352  0.8030725724064472
  0.7456778704537873  0.7498632225222099  0.8036276618908668
  0.5010572296398286  0.7432994347763617  0.8068746543969344
  0.5004113136271020  0.5004729333655571  0.7986256112347355
  0.7473663458278997  0.0032377095651933  0.0006765004633182
  0.7475364180124415  0.7537487184366078  0.0012867489065531
  0.7453535204788281  0.5021365797473939  0.9997518078925229
  0.9948664751873311  0.2518694721830108  0.9999339305313112
```

| | | |
|---|---|---|
| 0.5014171461749657 | 0.7519317426345515 | 0.2024025325167323 |
| 0.2504171453943702 | 0.7502323481678890 | 0.2011212955629877 |
| 0.5027316183992897 | 0.4973298982259918 | 0.2002897647953545 |
| 0.0022036879095925 | 0.0002145867070338 | 0.3972501686909277 |
| 0.2536012155556954 | 0.0030919533549330 | 0.3952288166071654 |
| 0.7528215049570818 | 0.7461226589563547 | 0.4066128047005398 |
| 0.7531177462927351 | 0.4997317469489121 | 0.3976047472233362 |
| 0.5017665116256134 | 0.2438207887760924 | 0.4017390969842769 |
| 0.5013106681580564 | 0.9898645713677189 | 0.6052227653775355 |
| 0.2497610500610229 | 0.5012258304789813 | 0.5974759879704471 |
| 0.4999778138171000 | 0.2503892423541072 | 0.7975603438962948 |
| 0.2463192388123607 | 0.2520073719438189 | 0.7969168969769327 |
| 0.9993963892930736 | 0.7525903654048797 | 0.0002429113647858 |
| 0.4964243213142321 | 0.7522519406781343 | 0.0046408446083932 |
| 0.4943168004103802 | 0.5030114948405120 | 0.0002961097053921 |
| 0.2490302739065983 | 0.5051960581127150 | 0.9963068253304351 |
| 0.2458815179554356 | 0.2559812389028734 | 0.9949391268414524 |
| 0.0015163050610363 | 0.9994046684422695 | 0.2032625631375623 |
| 0.7506357703972596 | 0.5005468447181547 | 0.1981491978991818 |
| 0.2502247050109274 | 0.2546397442978189 | 0.1925468608690811 |
| 0.0039311590521910 | 0.7465058153755458 | 0.4018622662781464 |
| 0.2526534077457545 | 0.7514158584719249 | 0.3981971834437622 |
| 0.2508790898766943 | 0.7526556858873504 | 0.5972307809761035 |
| 0.0052714921788748 | 0.4929455296623566 | 0.5977358647933450 |
| 0.7557039467640464 | 0.4916538712685150 | 0.5993378359790800 |
| 0.2514638932468352 | 0.2449445822403434 | 0.6009019770329304 |
| 0.9980036756604908 | 0.2505369696523883 | 0.7984626158375213 |
| 0.7480583685978457 | 0.2522953074059995 | 0.7976349517599139 |
| 0.0013567773802361 | 0.9998320317379181 | 0.0016492912725538 |
| 0.7489460266769723 | -0.0000003000121770 | 0.2036857868100210 |
| 0.9991782325398341 | 0.5063978962029239 | 0.1960249466333188 |
| 0.7543564710954563 | 0.2434350056618258 | 0.2037124209625082 |
| 0.5014358911785477 | 0.7497495544469919 | 0.4036772713555110 |
| 0.2509897893535531 | 0.4981640153968563 | 0.3997309879707929 |
| 0.0014682230789617 | 0.2547009284830993 | 0.3936577804856680 |
| 0.0029864924870785 | 0.0008854271799222 | 0.5986646438105849 |
| 0.7463681639936475 | 0.9994911462788371 | 0.6032683857484051 |
| 0.0003018233506070 | 0.7495700058342640 | 0.5998063946520269 |
| 0.7527482413661643 | 0.7388618023962762 | 0.6062715098880150 |
| 0.5038019032808123 | 0.4971012249645013 | 0.6001281274770320 |
| 0.5003152397326056 | 0.2496780635522006 | 0.5990605681746106 |
| 0.2488409130884029 | 0.9971471880025186 | 0.8035991109659693 |
| 0.9976484186341733 | 0.7537532941386419 | 0.8014737826398646 |
| 0.7515549282424008 | 0.4938596057837966 | 0.8002693246005715 |
| 0.8732399557225032 | 0.8785977107856441 | 0.0999128699514922 |
| 0.6261336004100904 | 0.8760636792544095 | 0.1093392740348446 |
| 0.3747911809264339 | 0.8786599836311472 | 0.1049237785062352 |
| 0.1278670477424974 | 0.8708435865850487 | 0.1038925182340930 |
| 0.8740008094351777 | 0.6266461066539725 | 0.0968701995100748 |
| 0.6241734264713241 | 0.6267942987327357 | 0.1002159088586600 |
| 0.3834107130064043 | 0.6189731123810338 | 0.1059965386667254 |
| 0.1252822373337073 | 0.6340670319683120 | 0.0937231831929670 |
| 0.8699786803053465 | 0.3824050260963920 | 0.0963820176820034 |
| 0.6223068118219480 | 0.3698263690084543 | 0.0958339119932262 |
| 0.3656205569294230 | 0.3859298689569188 | 0.0910280258416801 |
| 0.1291502109217263 | 0.3773651998758039 | 0.0948882858199206 |
| 0.8785393833483519 | 0.1176009955421986 | 0.1085913086102739 |
| 0.6178181807495681 | 0.1314419377124067 | 0.0999340717644033 |
| 0.3652058607038012 | 0.1338264429700659 | 0.0887005917405286 |
| 0.1171148253988295 | 0.1344787203798773 | 0.0932914702689699 |
| 0.8815599369881224 | 0.8709192652706615 | 0.3088348237105872 |
| 0.6274336288330858 | 0.8733043397516318 | 0.3126284692209049 |
| 0.3828584948776867 | 0.8727351229277499 | 0.3030982011357352 |
| 0.1295886484044980 | 0.8749006873534706 | 0.2994351326320576 |
| 0.8683052724984385 | 0.6337902392249390 | 0.2959917165066609 |
| 0.6286094618567230 | 0.6207542732850210 | 0.2985078061975937 |
| 0.3754689419460352 | 0.6262316515510847 | 0.3005074719138822 |
| 0.1241626074377744 | 0.6221226380480053 | 0.3016415960102353 |
| 0.8749066380552267 | 0.3776997300591723 | 0.2958098374964995 |
| 0.6256685333845086 | 0.3734343244897954 | 0.2975420616916356 |
| 0.3860783385047562 | 0.3639849217644286 | 0.2961194604713068 |
| 0.1275390355361058 | 0.3807255771832997 | 0.2955528764268330 |
| 0.8852303383823528 | 0.1270564232634102 | 0.2969911180944521 |
| 0.6308977617975039 | 0.1068458625841827 | 0.3153613268959002 |
| 0.3755674206149435 | 0.1345916825255849 | 0.2833579660502317 |
| 0.1340544072308520 | 0.1222216294715386 | 0.2922801231879442 |
| 0.8752849923599670 | 0.8760697830570163 | 0.4980039361349626 |
| 0.6246261898782636 | 0.8512612381307648 | 0.5198214154982456 |
| 0.3666229919491947 | 0.8875257088545733 | 0.4995887582728107 |
| 0.1268999641224710 | 0.8753341706092604 | 0.4975986381381450 |
| 0.8866575093867020 | 0.6144289423870760 | 0.4980306574532585 |
| 0.6287064630697430 | 0.6190160209975460 | 0.5024521996552662 |
| 0.3702330540948224 | 0.6301782322268846 | 0.4990169393826288 |
| 0.1267310992727841 | 0.6266091831922274 | 0.4972939860102011 |
| 0.8836534642487049 | 0.3841635829448108 | 0.4914966675627660 |
| 0.6291180789527060 | 0.3738485469208568 | 0.4950445474117227 |
| 0.3780617072523225 | 0.3706787866992924 | 0.4997258530782061 |
| 0.1387490815398739 | 0.3641147852551758 | 0.5039314544762782 |
| 0.8726601884005830 | 0.1279001196049090 | 0.5009623454040961 |
| 0.6281315033007067 | 0.1151070922316147 | 0.4984495066814504 |
| 0.3778453278175888 | 0.1141445498831846 | 0.5019158799799686 |
| 0.1220268347426818 | 0.1328901228344935 | 0.4895923630954519 |
| 0.8639621087138799 | 0.8755988841081853 | 0.7083115684636391 |
| 0.6243452827957462 | 0.8629461150855883 | 0.7139319873652772 |
| 0.3874037466140753 | 0.8589820494888696 | 0.7133584866651421 |
| 0.1328331131627438 | 0.8766730902374933 | 0.6939350516281174 |
| 0.8844799597052473 | 0.6154861064829004 | 0.6941118064286704 |
| 0.6350314480566694 | 0.5999909589110584 | 0.7039347398670111 |
| 0.3800867005994452 | 0.6274000530048865 | 0.6998313614759420 |
| 0.1271900144546802 | 0.6288850296439615 | 0.6979523428637736 |
| 0.8818474612637903 | 0.3658547501359297 | 0.6976707349301815 |
| 0.6419728645528197 | 0.3579869459831907 | 0.6912571816813884 |
| 0.3710901966923679 | 0.3807233982225984 | 0.6983954649970875 |
| 0.1246766246699404 | 0.3736438631604644 | 0.6982273605672024 |
| 0.8708548148379744 | 0.1189903660254990 | 0.7043617904812538 |
| 0.6127508738271286 | 0.1245034714538051 | 0.6999993831230609 |
| 0.3716003921908428 | 0.1229221558680281 | 0.6984379774500201 |
| 0.1272320847309349 | 0.1222145321811691 | 0.6946310197910507 |
| 0.8779787334694743 | 0.8712802141497428 | 0.9069778664536766 |
| 0.6139903819242920 | 0.8864535530801632 | 0.9115648286728776 |
| 0.3679576566206000 | 0.8769359592073889 | 0.9150639290484897 |
| 0.1147366706809585 | 0.8837118444524916 | 0.8960939601217629 |
| 0.8650464654628737 | 0.6314230404285855 | 0.8962046038338117 |
| 0.6188534781915817 | 0.6314382140518637 | 0.9026956741488621 |
| 0.3675975611750505 | 0.6361524516565411 | 0.9104086761848026 |
| 0.1249919937343120 | 0.6272816233226727 | 0.9007155718149699 |
| 0.8692229563143472 | 0.3727301645114298 | 0.9024068131211058 |
| 0.6112560558418049 | 0.3823064233284463 | 0.8952352176211149 |
| 0.3749600828153410 | 0.3839894652694650 | 0.8930061442259899 |
| 0.1192251007320968 | 0.3807383365533530 | 0.8977870184435562 |
| 0.8622809367460111 | 0.1383748830530895 | 0.8999792477688666 |
| 0.6236383147717327 | 0.1248568940699772 | 0.8986652697073098 |
| 0.3637163251145120 | 0.1391294414267818 | 0.8935856818632311 |
| 0.1165154683381999 | 0.1324030389377863 | 0.8967138491246350 |